\documentclass[prl,twocolumn,showpacs,preprintnumbers,amsmath,amssymb,floats]{revtex4-1}
\pdfoutput=1
\usepackage{graphicx}
\usepackage{pstricks}
\usepackage{amsmath}
\usepackage{color}
\usepackage{amsbsy}
\usepackage{lipsum}
\usepackage{hyperref}
\usepackage{dcolumn}
\usepackage{bm}
\usepackage{soul}
\newcommand{\mbf}{\mathbf} 
\renewcommand{\k}{{\mbf k}}

\begin{document}
\title{Model of Electronic Structure and Superconductivity in Orbitally Ordered FeSe}
\author{Shantanu Mukherjee$^1$, A. Kreisel$^1$, P. J. Hirschfeld$^2$, and Brian M. Andersen$^1$}
\affiliation{$^1$Niels Bohr Institute, University of Copenhagen, Universitetsparken 5, DK-2100 Copenhagen,
Denmark\\
$^2$Department of Physics, University of Florida, Gainesville, Florida 32611, USA}
\date{February 11, 2015}

\begin{abstract}

We provide a band structure with low-energy properties consistent with recent photoemission and quantum oscillations measurements on FeSe, assuming mean-field like $s$ and/or $d$-wave orbital ordering at the structural transition. We show how the resulting model provides a consistent explanation of the temperature dependence of the measured Knight shift and the spin-relaxation rate. Furthermore, the superconducting gap structure obtained from spin fluctuation theory exhibits nodes on the electron pockets, consistent with the 'V'-shaped density of states obtained by tunneling spectroscopy on this material, and the temperature dependence of the London penetration depth. Our studies prove that the recent experimental observations of the electronic properties of FeSe are consistent with orbital order, but leave open the microscopic origin of the unusual band structure of this material.

\end{abstract}

\pacs{71.18.+y, 74.20.Rp, 74.25.Jb, 74.70.Xa}
\maketitle

The electronic properties and the nature of the interactions that drive the low-energy physics and the ordered phases of iron-based superconductors (FeSC) continue to pose an outstanding problem in modern condensed matter physics. The diversity of the properties among the different families of FeSC and their complex multi-orbital band structure have hindered the understanding of the electronic states in these materials, as well as the mechanism of superconductivity.

A material that stands out is the structurally simplest compound, FeSe, which exhibits a tetragonal to orthorhombic structural phase transition at $T_S\sim 90$ K without concomitant spin density wave (SDW) order, and becomes superconducting below $T_c\sim 9\,\text{K}$. Below $T_S$, the material exhibits strong electronic anisotropy and the absence of tetragonal symmetry-breaking SDW order, makes FeSe ideal for studying the origin and consequences of nematicity, i.e. the breaking of rotational symmetry while preserving translational symmetry. For example, an early scanning tunneling microscopy (STM) study of FeSe films on SiC substrate found highly elongated vortices and
impurity states, and an associated nodal superconducting gap,\cite{song11} but until recently similar experiments on crystals were hampered
by sample quality. Other remarkable properties of FeSe include the significant enhancement of the superconducting critical temperature $T_c$ both under pressure,\cite{medvedev09} and for monolayers of FeSe grown on SrTiO$_3$ surfaces.\cite{yan12,tan13,ge14}

\begin{figure}[b]
\includegraphics[width=\columnwidth]{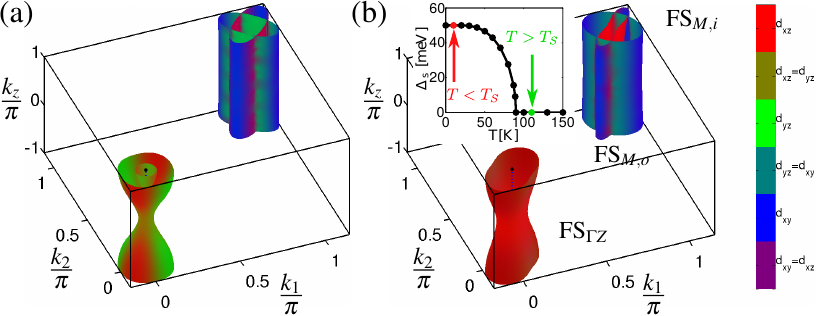}
\caption{(Color online) (a) Fermi surface (FS) above $T_S$ with SO coupling, and (b) FS at low $T$ with additional orbital splitting of $50\,\text{meV}$ consisting of a $\Gamma$ centered FS (FS$_{\Gamma Z}$) cylinder and an inner and outer FS centered around the M point (FS$_{M,i}$ and FS$_{M,o}$). The inset shows the $T$ dependence of the orbital splitting $\Delta_{\text{s}}(T)$; the two colored (gray) dots represent the $T$ chosen for the two displayed FS.}
\label{fig:FS}
\end{figure}

Recently, the study of bulk FeSe crystals has been revitalized by the growth of very clean samples amenable to study the details of the low-energy properties by e.g. nuclear magnetic resonance (NMR), transport, STM, angular resolved photoemission spectroscopy (ARPES), and quantum oscillation (QO) experiments.\cite{bohmer13} Even though a consensus on the electronic bands has not yet been reached by ARPES,\cite{nakayama14,maletz14,shimojima14,watson14,ZhangP15,SuzukiY15,ZhangY15} recent studies found that the Fermi surface (FS) above $T_S$ consists of two small hole cylinders of mainly $d_{xz}/d_{yz}$ character around the $\Gamma-Z$ line. The hole bands are split by a sizable spin-orbit coupling (SO) of $\lambda_{\text{Fe}}\simeq 20\,\text{meV}$ above the structural transition, and turns into a single elongated hole cylinder below $T_S$.\cite{watson14} ARPES also finds an electron pocket at the $M$ point of mainly $d_{xz}/d_{yz}$ character. Importantly, the expected $d_{xz}$/$d_{
yz}$ degeneracy at $M$ is lifted by $\sim 50\,\text{meV}$, constituting strong evidence for orbital order in FeSe. We emphasize that these results for the electronic structure are very different from those obtained within DFT calculations.\cite{maletz14,Eschrig09} For example, ARPES finds that the electronic bands in FeSe are  renormalized compared to DFT calculations by a factor of $\sim$3 for the $d_{xz}/d_{yz}$ bands and $\sim$9 for the $d_{xy}$ band.\cite{maletz14,watson14} QO performed at low $T$ in magnetic fields large enough to suppress superconductivity are consistent with the ARPES data in observing small largely 2D pockets, even though the amount of dispersion along $k_z$ remains unsettled.\cite{terashima14,watson14}

Recent $^{77}$Se NMR measurements on FeSe have reported a clear splitting of the NMR line shape setting in at $T_S$, with an order parameter-like $T$ dependence below $T_S$.\cite{baek14} At high $T$, the spin-lattice relaxation rate is, however, unaffected by the structural transition, and only exhibits a clear upturn at low $T$ closer to the superconducting $T_c$.\cite{baek14,bohmer15} These recent experiments have been interpreted as evidence for orbitally
driven nematic behavior in FeSe, but despite the apparent weakness of (momentum summed) spin fluctuations near $T_S$, the spin-nematic
picture may still apply.  One possibility is that, unlike other FeSC, fluctuations in FeSe with different wave vectors compete to frustrate long-range magnetic
order.\cite{glasbrenner15}

Finally, we note that the resulting Fermi energy of the bands of FeSe seen by ARPES are remarkably small, comparable to the superconducting gap, which suggests the possibility that FeSe may be close to a BEC/BCS crossover, and thus exhibit unusual thermodynamic properties and magnetic field effects.\cite{Kasahara14}  Thus, for multiple reasons, it is important to perform new theoretical studies of this intriguing material and obtain a minimal model capturing its main electronic properties.

Here, we perform a theoretical study of the consequences of orbital order in a band relevant to FeSe. Starting from the DFT-generated band for FeSe obtained by Eschrig \textit{et al.}\cite{Eschrig09}, we apply band renormalization of $H_{TB}=H_{0}/z$ (where $z=6$ is renormalization factor and $H_{0}$ is the unrenormalized tight-binding Hamiltonian) and additional shifts to the hopping integrals (see Supplementary Information (SI) for details) to generate a new tight-binding model. We find that a band consistent with ARPES and QO is only possible in the presence of an orbital splitting setting in at $T_S$, and a $T$ independent SO coupling. Further, we explain the recent Knight shift and the spin relaxation rate measurements, and study how spin fluctuation-mediated pairing can lead to a nodal gap structure in agreement with measured density of states (DOS) and penetration depth $\lambda$ of FeSe.

\begin{figure}[b]
\includegraphics[width=\columnwidth]{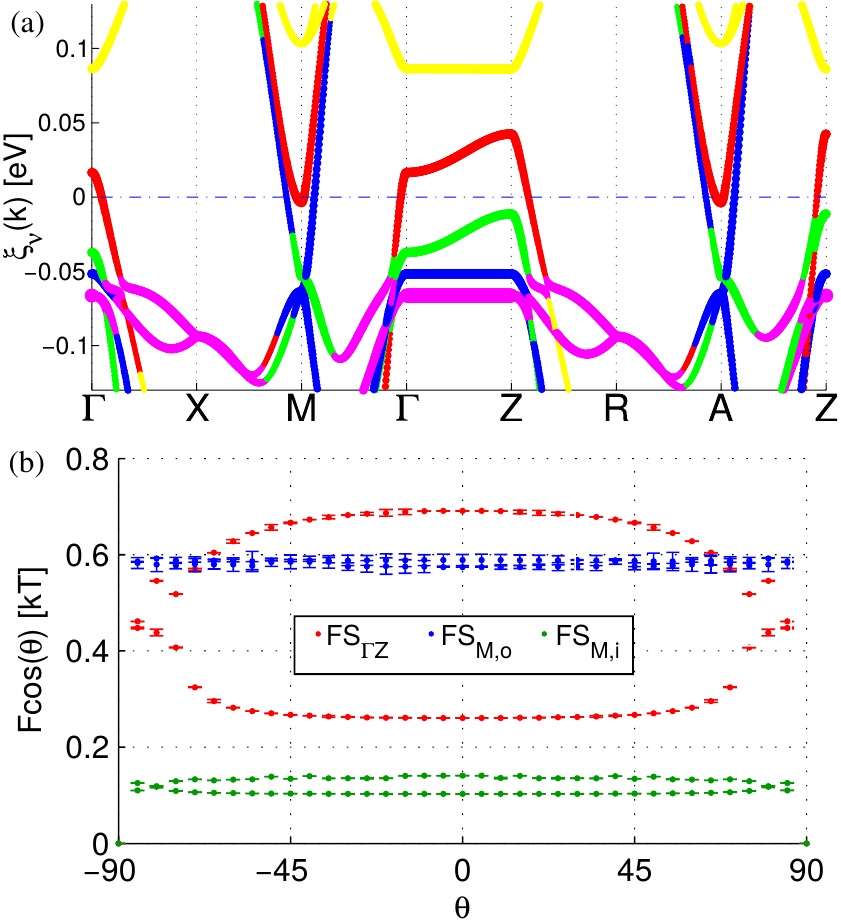}
\caption{(Color online) (a) Band structure of the 10 orbital model at low $T$ with orbital order and SO coupling, yielding the QO frequencies as a function of magnetic field angle $\theta$ shown in (b), where the error bars indicate the numerical uncertainty in the determination of the extremal orbits. The orbital character in the band plot is indicated by the colors red $d_{xz}$, green $d_{yz}$, blue $d_{xy}$, yellow $d_{x^2-y^2}$, purple  $d_{3z^2-r^2}$.
}
\label{fig:QO}
\end{figure}

The bare Hamiltonian used in this study is given by
\begin{eqnarray}
H&=&H_{TB} + H_{OO},\\
H_{TB}&=&\sum_{\bf{k},\mu,\nu,\sigma}t_{\mu\nu}({\bf{k}})c^{\dag}_{\mu\sigma}({\bf{k}})c_{\nu\sigma}({\bf{k}}),\\
H_{OO}&=&\Delta_{s}(T)\sum_{{\bf{k}}\sigma}(n_{xz\sigma}({\bf{k}})-n_{yz\sigma}({\bf{k}})).
\end{eqnarray}
Here ($\mu,\nu$) are orbital indices, $t_{\mu\nu}({\bf{k}})$ are the hopping integrals, and $n_{\mu\sigma}({\bf{k}})=c^{\dag}_{\mu\sigma}({\bf{k}})c_{\mu\sigma}({\bf{k}})$. All details of the hopping integrals are provided in the SI for both a 5-orbital and 10-orbital model. In the orbitally ordered state, $H_{OO}$ contributes and $\Delta_{s}(T)$ is assumed to exhibit a mean-field $T$ dependence with a maximum amplitude $\Delta_{s}(T=0)=50\,\text{meV}$. For simplicity we focus in the main part of this paper on a pure $s$-wave OO, but the consequences of an additional $d$-wave OO of the form $\Delta_{d}(T)\sum_{{\bf{k}}\sigma}(\cos(k_x)-\cos(k_y))(n_{xz\sigma}({\bf{k}})+n_{yz\sigma}({\bf{k}}))$ have also been studied and the results can be found in the SI. It has been reported by ARPES that the band splitting of the $d_{xz}/d_{yz}$ bands at the M-point in the orbitally ordered state does indeed show a mean-field behavior and saturates at low $T$ with a band splitting of $\sim 50\,\text{meV}$.\cite{nakayama14,shimojima14,ZhangP15,ZhangY15}. Finally, we have included a SO term, $H_{SO}=\lambda_{\text{Fe}}\sum_i\sum_{x,y,z}L^{\alpha}_iS^{\alpha}_i$, which causes a band splitting of $20\,\text{meV}$ in the tetragonal high $T$ phase.\cite{Friedel64,Kreisel13}

\textit{Band Structure.}
As shown in Fig.~\ref{fig:FS} and Fig.~\ref{fig:QO}, the band structure and the resulting FS of $H$ is in nearly quantitative agreement with experiments. Below $T_S$, the hole band at the $\Gamma$ point is split by a $50\,\text{meV}$ orbital order (at $T=10\,\text{K}$) and the bottom of the band lies $\sim20\,\text{meV}$ below the chemical potential. Similarly a dispersionless $d_{xy}$-band is present at an energy of $\sim-50\,\text{meV}$ at the $\Gamma$ point. At the M point, the electron pockets consist of quasi-2D cylinders where the outer pocket, having a dominant $d_{xy}$ character, encloses an inner $d_{xz}/d_{yz}$ electron pocket. The inner electron band at the M point has an orbital splitting of $50\,\text{meV}$ and almost grazes the Fermi level. These low $T$ band structure values are in good agreement with ARPES results.\cite{watson14,nakayama14,shimojima14,ZhangP15,ZhangY15}

Similar agreement with ARPES is achieved for $T>T_S$ where the orbital order is absent. There, the hole pockets consist of a quasi-2D outer circular cylinder and an inner hole pocket near the Z-point as seen in Fig.~\ref{fig:FS}(a). The band also exhibits overall agreement with the orbital content observed in polarized ARPES experiments.\cite{watson14} At high $T$, the hole pocket at the zone center and the inner electron pocket at the M point contain both $d_{xz}$ and $d_{yz}$ character, and the outer electron pocket is predominantly of $d_{xy}$ character. Similar orbital content of the Fermi pockets have also been seen in ARPES measurements.\cite{shimojima14,watson14,ZhangP15,SuzukiY15} At low $T$, the orbital content of the hole cylinder is dominated by $d_{xz}$ character ($d_{yz}$ for the other twin). For pure $s$-wave orbital order, both the hole and the inner electron cylinders have dominant $d_{xz}$ orbital character (see Fig.~\ref{fig:FS}(b)) whereas in the presence of an additional $d$-wave orbital order the electron cylinder can have the opposite $d_{yz}$ orbital character (see SI). For the electron pocket at low $T$, the inner pocket contains both $d_{xz}$ and $d_{yz}$ orbital character whereas the outer pocket at low $T$ has orbital content dominated by the $d_{xy}$-orbital. Although the orbital content of the electron pockets agree well with experiments,\cite{watson14,ZhangP15,SuzukiY15} the outer $d_{xy}$ electron pocket is difficult to observe in ARPES due to matrix elements effects.

\textit{Quantum Oscillations.}
The extremal FS areas in FeSe at low $T$ as well as their $k_z$ dispersion have been studied by QO measurements.\cite{watson14,terashima14,audouard14} These experiments have found four well separated QO frequencies, and arguments have been put forward that the QO frequencies
correspond to one electron and one hole quasi-2D FS cylinder,\cite{terashima14} as well the possibility of a single quasi-2D hole cylinder and two almost dispersionless electron cylinders.\cite{watson14} Although the former possibility cannot be ruled out, in this study we have pursued the latter possibility, which is supported by the weak $k_z$ dispersion observed for the electron cylinders.\cite{watson14}

Starting from the 10 orbital tight-binding Hamiltonian and including the effects of SO coupling, we calculate the eigenenergies $\xi_i({\bf{k}})$ on a grid in the Brillouin zone (BZ) and obtain the extremal areas $F$ of the FS for cuts on planes perpendicular to the external magnetic field using a numerical method.\cite{Rourke12,Diehl14} The direction of the magnetic field is then parametrized by the angle $\theta$ between the crystallographic $c$ axis and the field direction. For $\theta=0$, the electron pockets have extremal areas of $F\sim 588\,\text{T}$ for the $d_{xy}$ pocket and $F\sim 102\,\text{T}$ for the smaller $d_{xz}/d_{yz}$ pocket, as seen from Fig.~\ref{fig:QO}(b). The hole Fermi cylinder is elongated due to the effect of orbital ordering with a maximum area of $F\sim691\,\text{T}$ for $k_z=\pi$ and a minimum area of $F\sim260\,\text{T}$ at $k_z=0$. Overall the experimentally observed binding energies for the $d_{xz}/d_{yz}$ and $d_{xy}$ bands, the 3D FS structure of both hole and electron
pockets, the extremal orbit areas as well as their $k_z$ dispersion are in good agreement with our calculations. We have also calculated the Sommerfeld coefficient from the effective masses extracted from our quantum oscillation calculation. Using the recipe given in Ref.~\onlinecite{watson14}, we find a Sommerfeld coefficient of 4.5 mJ/mol-K$^2$ which is in reasonable agreement with experimental value of $\sim$5.3-5.7 mJ/mol-K$^2$.\cite{bohmer15,lin11,hafiez15}

\textit{Nuclear Magnetic Resonance.}
Next, we test our electronic model of FeSe to see if it can also reproduce NMR experiments.\cite{baek14,bohmer15} For computational simplicity, in the following we apply the 5 orbital model that shows a good agreement with the 10 orbital model (see SI) and ignore the effect of SO coupling which causes only small quantitative changes to the observables discussed in the remainder of this paper. The NMR Knight shift is proportional to the homogeneous susceptibility, $K=A_{hf}\chi_{RPA}({\bf{q}}=0)+K_{\text{chem}}$, where we have approximated the spin susceptibility by its standard RPA form, $A_{hf}$ is the hyperfine form factor, and $K_{\text{chem}}$ is a $T$ independent chemical shift which we have ignored for the purposes of this study. For the following calculations, we include local Coulomb interactions via the standard Hubbard-Hund Hamiltonian\cite{Kuroki08} parametrized by the Hubbard interaction $Uz$ ($z$ is the band renormalization) and the Hund's exchange $J$ (see SI), calculate the orbitally resolved non-interacting susceptibility, and include interactions within RPA.\cite{s_graser_09}

In the paramagnetic state, the form factor is a diagonal matrix with components $(A^{xx}_{hf},A^{yy}_{hf},A^{zz}_{hf})$ where the coordinates point along the Fe-Fe direction representing the magnetic field orientation of the NMR experiment. The form factor maintains the symmetry of the underlying lattice such that for the high $T$ tetragonal phase $A^{xx}_{hf}=A^{yy}_{hf}\neq A^{zz}_{hf}$ whereas the orbital ordered orthorhombic phase has $A^{xx}_{hf}\neq A^{yy}_{hf}\neq A^{zz}_{hf}$. This anisotropy leads to a split Knight shift frequency below $T_S$ in twinned samples of FeSe.\cite{baek14,bohmer15}

\begin{figure}[b]
\includegraphics[width=\columnwidth]{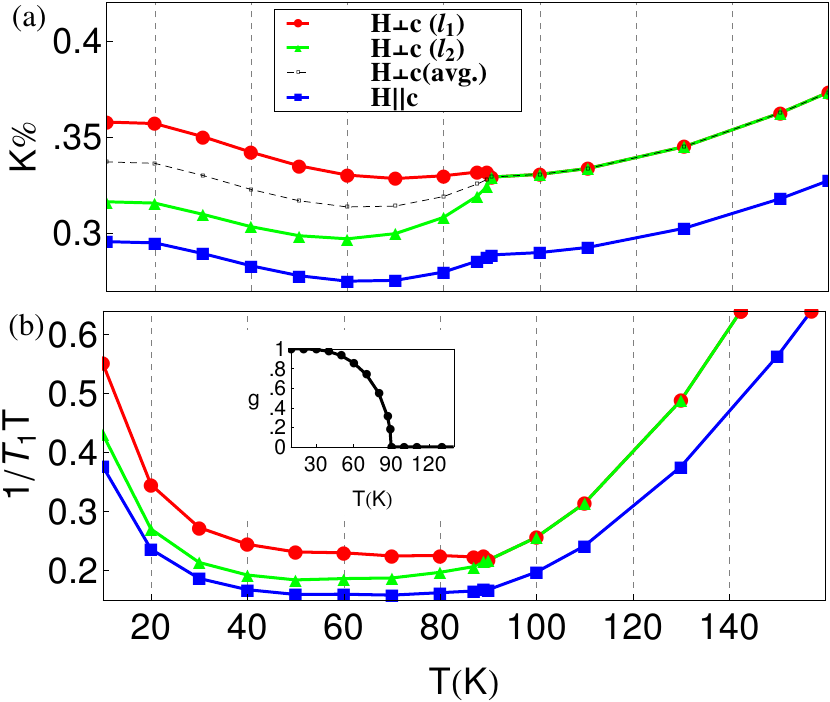}
\caption{(Color online) (a) NMR Knight shift versus $T$. The hyperfine form factor has been taken as $A^{l_1/l_2}_{hf}=0.6[0.57\pm0.035g(T)]$ for $H\perp c$ and $A^c_{hf}=0.6\times0.5$ for $H || c$. (b) Spin-lattice relaxation rate versus $T$ with $A^{l_1/l_2}_{hf}=0.57\pm0.035g(T)$ for $H \perp c$ and $A_{hf}=0.5$ for $H || c$. Red curve $H \perp c$ (domain $l_1$), Green curve $H \perp c$ (domain $l_2$), Black curve $H \perp c$ (domain average), Blue curve $H || c$.}
\label{fig:knight}
\end{figure}

As shown by Baek {\it{et al.}},\cite{baek14} the Knight shift splitting exhibits a $T$ dependence proportional to the mean-field orbital order parameter. Therefore, we model the form factor by the expression $A_{hf}=\alpha\pm g(T)$, where $g(T)=\beta\Delta_{s}(T)$, $(\alpha,\beta)$ are fitting parameters, and $\pm$ refers to the two orthorhombic domains $l_1$ and $l_2$. The calculated Knight shift as a function of $T$ is shown in Fig.~\ref{fig:knight}(a). At high $T$ above $T_S$, the Knight shift increases with $T$ similar to experiments, in contrast to the DFT generated non-renormalized bands (see SI). Below $T_S$, for a particular magnetic field direction the Knight shift shows a minimum value around $T\sim60\,\text{K}$ similar to experimental results. Below $T\sim60\,\text{K}$, we find a slight enhancement of the Knight shift signal. Although the measured Knight shift  saturates
and does not show this enhancement for both orthorhombic domains, this may be simply related to a deviation of the splitting from
mean-field behavior found experimentally at the lowest $T$.\cite{baek14}

In order to study the evolution of the spin fluctuations, we have also calculated the spin-lattice relaxation rate
\begin{align}
	\frac{1}{T_1T}=\lim_{\omega_0\rightarrow 0}\frac{\gamma_N^2}{2N}k_B\sum_{{\bf{q}} \xi \psi} |A^{\xi\psi}_{hf}({\bf{q}})|^2
\frac{\mathop\text{Im} \{\chi^{\xi \psi}_{RPA}({\bf{q}},\omega_0)\}}{ \hbar \omega_0}.
\end{align}
NMR experiments probing the $^{77}$Se atoms in FeSe exhibit a $q$ dependent hyperfine form factor in the paramagnetic state given by $A^{\xi\psi}_{hf}({\bf{q}})=A^{\xi\psi}_{hf}\cos(q_x/2)\cos(q_y/2)$ assuming that the Se ion interacts with its four nearest Fe neighbors only. Since $^{77}$Se is a spin 1/2 ion, quadrupole type coupling to local lattice distortions do not contribute to the relaxation rate. As seen from the form factor, spin fluctuations at the edges of the BZ will be filtered out. The result of the calculation for $1/{T_1T}$ for interaction parameters $Uz=1.8\,\text{eV}$ and $Jz=0.1Uz$ is shown in Fig.~\ref{fig:knight}(b). As seen, the spin fluctuations are enhanced at low $T$. However, as observed in recent NMR experiments\cite{baek14,bohmer15} the enhancement does not occur at $T_S$ despite the sharp increase of $\Delta_{s}$ at $T_S$, but below about $T\sim40\,\text{K}$. Interestingly, this increase of spin fluctuations at low $T$ is caused by the orbital ordering which leads to a low-$T$
incommensurability in the spin susceptibility that pushes spectral weight away from BZ edges, and therefore does not allow the structure factor to effectively filter out those fluctuations (see SI). Note that although the low $T$ spin susceptibility avoids the magnetic state by remaining below the Stoner limit, the enhanced fluctuations at low $T$ have important consequences for spin-fluctuation mediated pairing.

\textit{Spin-fluctuation pairing.}
 \begin{figure}[t]
\includegraphics[width=\columnwidth]{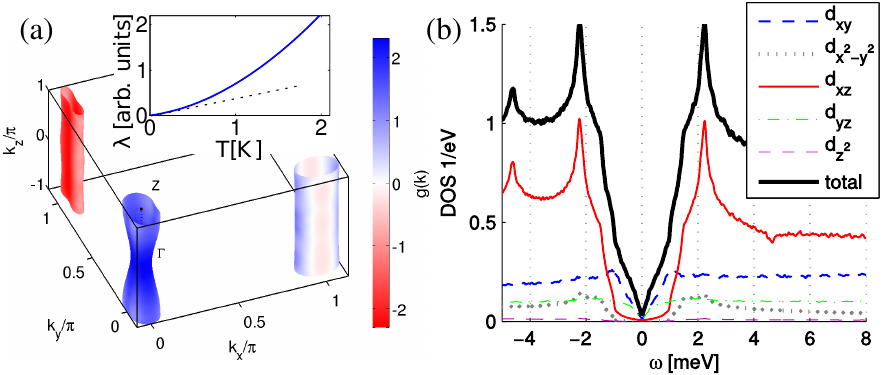}
\caption{(Color online) Superconducting order parameter as calculated from spin-fluctuation pairing using the interactions $Uz=1.8\,\text{eV}$ and $Jz=0.1Uz$ shows nodal regions on one of the electron pockets (a). The corresponding DOS clearly exhibits nodal behavior (b) and the penetration depth stays linear down to low $T$ in agreement with experiments (a, inset).}
\label{fig_pair}
\end{figure}
What is the dominant pairing instability for the low-$T$ orbitally ordered state? To answer this question, we consider the scattering vertex in the singlet channel projected onto the band space $\Gamma (\k,\k')$ (see SI) and solve the linearized gap equation
 \begin{equation}\label{eqn:gapeqn}
-\frac{1}{V_G}  \sum_j\int_{\text{FS}_j}dS'\; \Gamma(\k,\k') \frac{ g_\alpha(\k')}{|v_{\text{F}j}(\k')|}=\lambda_\alpha g_{\alpha}(\k),
 \end{equation}
 where $v_{\text{F}j}(\k')$ is the Fermi velocity of band $j$ and the integration is performed over $\text{FS}_j$ to obtain the gap symmetry functions $g_{\alpha}(\k)$ and the set of eigenvalues $\lambda_\alpha$. The largest eigenvalue corresponds to the leading instability and the corresponding eigenfunction determines the structure of the superconducting gap  $\Delta(\k)\sim g(\k)$ close to $T_c$. In order to solve Eq.~(\ref{eqn:gapeqn}), the FS is discretized using a Delaunay triangulation\cite{Kreisel13} such that it reduces to solving a matrix eigenvalue problem. In the absence of orbital order, the leading instability is $d$-wave with nodes on the hole pockets, and no accidental nodes on the electron pockets, whereas in the absence of any band renormalization the leading instability is a nodeless sign changing $s\pm$ state. In Fig.~\ref{fig_pair}(a) we show the result for the gap structure in the low-$T$ phase with orbital order. The character of the gap structure cannot be classified in $s$ or $d$-wave symmetry because the underlying band structure is only $C_2$ symmetric.\cite{kang14} As seen Fig.~\ref{fig_pair}(a)
the orbital order has strong effects on the position of the nodes, i.e. it removes the nodes from the hole pockets and induces nodal lines on the $X$-centered ($Y$ for the other twin) electron pocket. The associated DOS (maximum gap set to $\approx 2.2\,\text{meV}$) and the linear-$T$ behavior of the low-$T$ penetration depth $\lambda$ shown in Fig.~\ref{fig_pair} are remarkably similar to recent experimental findings.\cite{song11,Kasahara14}

In summary, we have presented a model for the electronic structure of FeSe that includes orbital ordering, which is consistent with recent ARPES and QO experiments on high quality FeSe samples. This band, along with the standard local interaction potentials and exchanges,
explains both the $T$ dependence of the NMR Knight shifts and spin-relaxation rate, and leads to a pairing state with nodes and a $T$ dependence of the London penetration depth in agreement with a series of recent experiments on FeSe.

We acknowledge useful discussions with B. B\"{u}chner,  A. B\"ohmer, A. Coldea, T. Hanaguri, F. Hardy, M. Lang, I. I. Mazin, C. Meingast, T. Shibauchi, and R Valenti. We thank D. Guterding for using his QO code 
\href{https://github.com/danielguterding/dhva}{https://github.com/danielguterding/dhva}
A. K. and B. M. A. acknowledge financial support from a Lundbeckfond fellowship (grant A9318). P. J. H. was partially supported by US DOE DEFG02-05ER46236. This research
was supported in part by KITP under NSF grant PHY11-25915.

%

\newpage
\section{Supplementary Information}

\allowdisplaybreaks

\setcounter{equation}{0}
 \renewcommand{\thefigure}{\Roman{figure}}
  \renewcommand{\theequation}{S\arabic{equation}}
  \renewcommand{\topfraction}{0.95}
\renewcommand{\bottomfraction}{0.95}
\textit{Tight-Binding Bandstructure.} Here we provide additional information regarding the tight-binding parameters used for constructing the ten orbital model for FeSe. The tight-binding Hamiltonian for FeSe can be written as
\begin{eqnarray}
H_{TB}&=&\sum_{\bf{k},\mu,\nu,\sigma}t_{\mu\nu}({\bf{k}})c^{\dag}_{\mu\sigma}({\bf{k}})c_{\nu\sigma}({\bf{k}}).
\end{eqnarray}
The orbitals $\mu,\nu$ follow the same notation convention used in Eschrig {\it et al.} In particular, the ten orbitals are $[(xy)^{+},(x^2-y^2)^{+},(ixz)^{+},(iyz)^{+},(z^2)^{+},(xy)^{-},(x^2-y^2)^{-},(-ixz)^{-},(-iyz)^{-},(z^2)^{-}]$, where the + and - refer to the two iron atoms in the unit cell. To reproduce the experimental results we renormalize the Hamiltonian given in Ref.\onlinecite{Eschrig09} by a factor of 6 and include additional band shifts that lead to similar Fermi energies and Fermi pockets with shape and sizes similar to results of ARPES and quantum oscillation experiments. In general the Fermi surfaces seen in ARPES and QO experiments are best reproduced by including an additional spin orbit coupling of $\Delta_{OO}\sim20\,\text{meV}$ as shown in Fig.1 of the manuscript. The corresponding case without the SO coupling is shown in Fig.\ref{fig:FS_no_SO} below. One important difference between the two cases is the hole pockets at the zone center shown in  Fig.\ref{fig:FS_no_SO} become an outer hole cylinder when the effect of SO coupling is included in the tight binding model.  Below we list the band dispersion given in Ref.\onlinecite{Eschrig09} (with minor printing errors corrected).
\begin{gather*}
    H^{++}_{11}=\epsilon_1+2t^{11}_{11}(\cos k_1+\cos k_2 )\\
+2t^{20}_{11}[\cos(2k_1)+\cos(2k_2)]\\
+[2t^{001}_{11}+4t^{111}_{11}(\cos k_1 +\cos k_2 )\\
+4t^{201}_{11}(\cos 2k_x+\cos 2k_y )]\cos k_z ,\\
\\
    H^{++}_{12}=0,\\
\\
    H^{++}_{13}=2it^{11}_{13}(\sin k_1 -\sin k_2 )-4t^{201}_{14}\sin(2k_y)\sin k_z,\\
\\
    H^{++}_{14}=2it^{11}_{13}(\sin k_1+\sin k_2)-4t^{201}_{14}\sin(2k_x)\sin k_z,\\
\\
    H^{++}_{15}=2t^{11}_{15}(\cos k_1-\cos k_2),\\
\\
    H^{++}_{22}=\epsilon_2+2t^{11}_{22}(\cos k_1+\cos k_2),\\
\\
    H^{++}_{23}=2it^{11}_{23}(\sin k_1+\sin k_2),\\
\\
    H^{++}_{24}=2it^{11}_{23}(-\sin k_1 +\sin k_2),\\
\\
    H^{++}_{25}=0,\\
\\
    H^{++}_{33}=\epsilon_3+2t^{11}_{33}(\cos k_1 +\cos k_2 )\\
+2t^{20}_{33}\cos(2k_x)+2t^{02}_{33}\cos(2k_y)+4t^{22}_{33}\cos(2k_x)\cos(2k_y)\\
+[2t^{001}_{33}+4t^{201}_{33}\cos(2k_x)+4t^{021}_{33}\cos(2k_y)]\cos k_z,\\
\\
    H^{++}_{34}=2t^{11}_{34}(\cos k_1 -\cos k_2),\\
\\
   H^{++}_{35}=2it^{11}_{35}(\sin k_1 +\sin k_2),\\
\\
    H^{++}_{44}=\epsilon_3+2t^{11}_{33}(\cos k_1 +\cos k_2)+2t^{02}_{33}\cos(2k_x)\\
+2t^{20}_{33}\cos(2k_y)+4t^{22}_{33}\cos(2k_x)\cos(2k_y),\\
+[2t^{001}_{33}+4t^{021}_{33}\cos(2k_x)+4t^{201}_{33}\cos(2k_y)]\cos k_z,\\
\\
    H^{++}_{45}=2it^{11}_{35}(\sin k_1 -\sin k_2),\\
\\
    H^{++}_{55}=\epsilon_5\\
\\
    H^{+-}_{16}=2t^{10}_{16}(\cos k_x+\cos k_y)\\
+2t^{21}_{16}[(\cos k_1+\cos k_2)(\cos k_x+\cos k_y)\\
-\sin k_1(\sin k_x+\sin k_y)+\sin k_2(\sin k_x-\sin k_y)]\\
+4t^{101}_{16}(\cos k_x+\cos k_y)\cos k_z\\
+2t^{121}_{16}\{[\cos(k_1+k_y)+\cos(k_1+k_x)]\exp(ik_z)\\
+[\cos(k_2+k_y)+\cos(k_2-k_x)]\exp(-ik_z)\}\\
\\
    H^{+-}_{17}=0\\
\\
    H^{+-}_{18}=2it^{10}_{18}\sin k_x-4(t^{101}_{18}\sin k_x+t^{101}_{19}\sin k_y)\sin k_z\\
+2it^{121}_{19}[\sin(k_1+k_y)\exp(ik_z)-\sin(k_2+k_y)\exp(-ik_z)],\\
\\
    H^{+-}_{19}=2it^{10}_{18}\sin k_y-4(t^{101}_{19}\sin k_x+t^{101}_{18}\sin k_y)\sin k_z\\
+2it^{121}_{19}[\sin(k_1+k_x)\exp(ik_z)+\sin(k_2-k_x)\exp(-ik_z)],\\
\\
   H^{+-}_{1,10}=0\\
\\
   H^{+-}_{27}=2t^{10}_{27}(\cos k_x + \cos k_y),\\
\\
   H^{+-}_{28}=-2it^{10}_{29}\sin k_y,\\
\\
   H^{+-}_{29}=2it^{10}_{29}\sin k_x,\\
\\
   H^{+-}_{2,10}=2t^{10}_{2,10}(\cos k_x - \cos k_y),\\
\\
    H^{+-}_{38}=2t^{10}_{38}\cos k_x+2t^{10}_{49}\cos k_y\\
+2t^{21}_{38}[(\cos k_1+\cos k_2)\cos k_x-(\sin k_1-\sin k_2)\sin k_x]\\
+2t^{21}_{49}[(\cos k_1+\cos k_2)\cos k_y-(\sin k_1+\sin k_2)\sin k_y]\\
+4(t^{101}_{38}\cos k_x+t^{101}_{49}\cos k_y)\cos k_z\\
+2t^{121}_{38}[\cos(k_1+k_x)\exp(ik_z)+\cos(k_2-k_x)\exp(-ik_z)]\\
+2t^{121}_{49}[\cos(k_1+k_y)\exp(ik_z)+\cos(k_2+k_y)\exp(-ik_z)],\\
\\
    H^{+-}_{39}=4it^{101}_{39}(\cos k_x+\cos k_y)\sin k_z,\\
\\
    H^{+-}_{3,10}=2it^{10}_{4,10}\sin k_y,\\
\\
    H^{+-}_{49}=2t^{10}_{49}\cos(k_x)+2t^{10}_{38}\cos(k_y)\\
+2t^{21}_{49}[(\cos k_1+\cos k_2)\cos k_x-(\sin k_1-\sin k_2)\sin k_x]\\
+2t^{21}_{38}[(\cos k_1+\cos k_2)\cos k_y-(\sin k_1+\sin k_2)\sin k_y]\\
+4(t^{101}_{49}\cos k_x+t^{101}_{38}\cos k_y)\cos k_z\\
+2t^{121}_{49}[\cos(k_1+k_x)\exp(ik_z)+\cos(k_2-k_x)\exp(-ik_z)]\\
+2t^{121}_{38}[\cos(k_1+k_y)\exp(ik_z)+\cos(k_2+k_y)\exp(-ik_z)],\\
\\
     H^{+-}_{4,10}=2it^{10}_{4,10}\sin k_x,\\
\\
    H^{+-}_{5,10}=0.
\end{gather*}
Below we provide the tight-binding hoppings that fit the DFT band structure as given in Eschrig {\it et al}.\cite{Eschrig09}
The hopping shifts performed in this study have been written as an added term to the original hopping integrals that undergo a renormalization by a factor of $z=6$.
For the 2D dispersion the hopping integrals are
\begin{figure}[t]
\quad
\includegraphics[width=0.38\columnwidth]{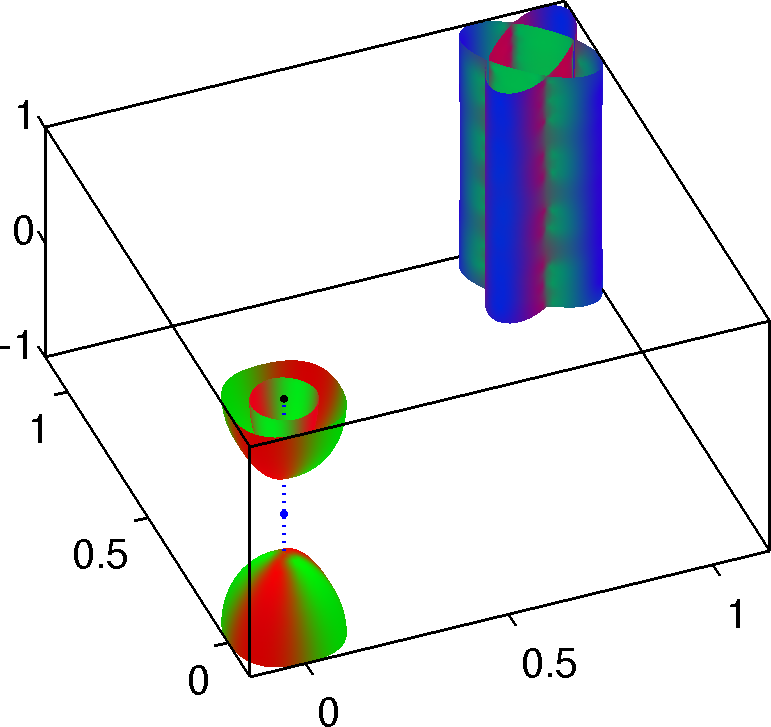}
\rput[tr](-0.38\columnwidth,0.36\columnwidth){(a)}
\rput[tr](-0.39\columnwidth,0.27\columnwidth){$\frac{k_z}\pi$}
\rput[tr](-0.33\columnwidth,0.07\columnwidth){$\frac{k_2}\pi$}
\rput[tr](-0.06\columnwidth,0.04\columnwidth){$\frac{k_1}\pi$}
\quad
\includegraphics[width=0.5\columnwidth]{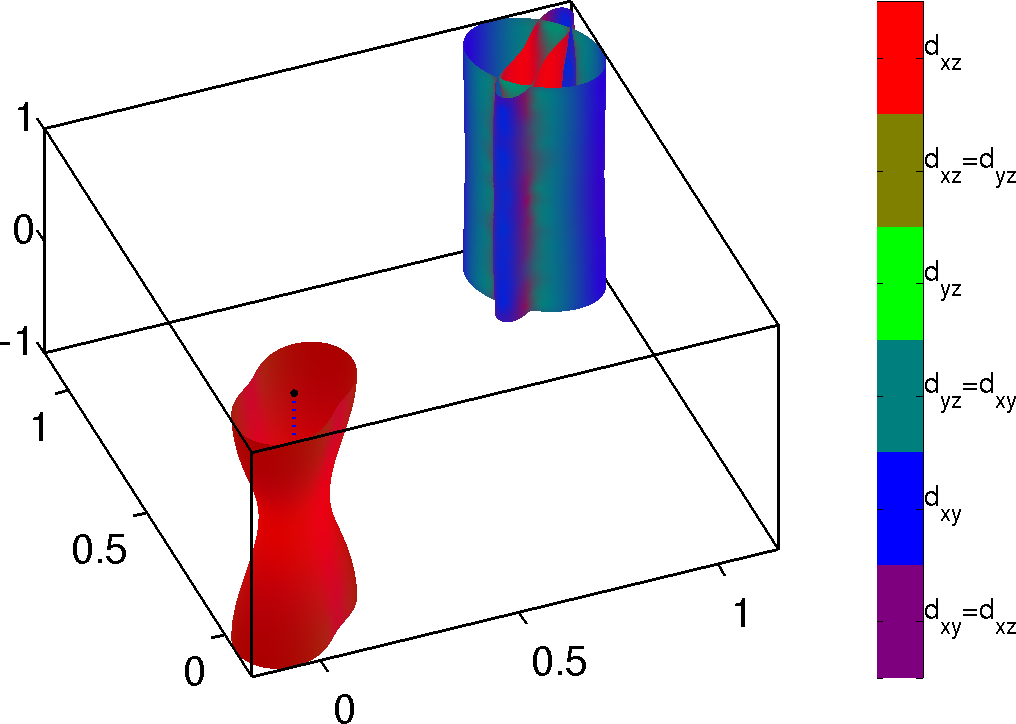}
\rput[tr](-0.51\columnwidth,0.36\columnwidth){(b)}
\rput[tr](-0.51\columnwidth,0.27\columnwidth){$\frac{k_z}\pi$}
\rput[tr](-0.45\columnwidth,0.07\columnwidth){$\frac{k_2}\pi$}
\rput[tr](-0.18\columnwidth,0.04\columnwidth){$\frac{k_1}\pi$}
\rput[tr](-0.120\columnwidth,0.35\columnwidth){\scriptsize {FS$_{M,i}$}}
\rput[tr](-0.20\columnwidth,0.19\columnwidth){\scriptsize {FS$_{M,o}$}}
\rput[tr](-0.25\columnwidth,0.1\columnwidth){\scriptsize {FS$_{\Gamma Z}$}}
\caption{(Color online) Fermi surface without SO coupling for $T>T_S$ (a) and $T<T_S$ (b). As in Fig. 1 of the main text, the Fermi surfaces are derived from the ten orbital model, plotted in the crystallographic BZ and visualized with the summed-color method where the absolute value of the overlap is mapped to the RGB value of the color on the surface as indicated by the color bar.
}\label{fig:FS_no_SO}
\end{figure}
\begin{figure}[t]
\includegraphics[width=0.98\columnwidth]{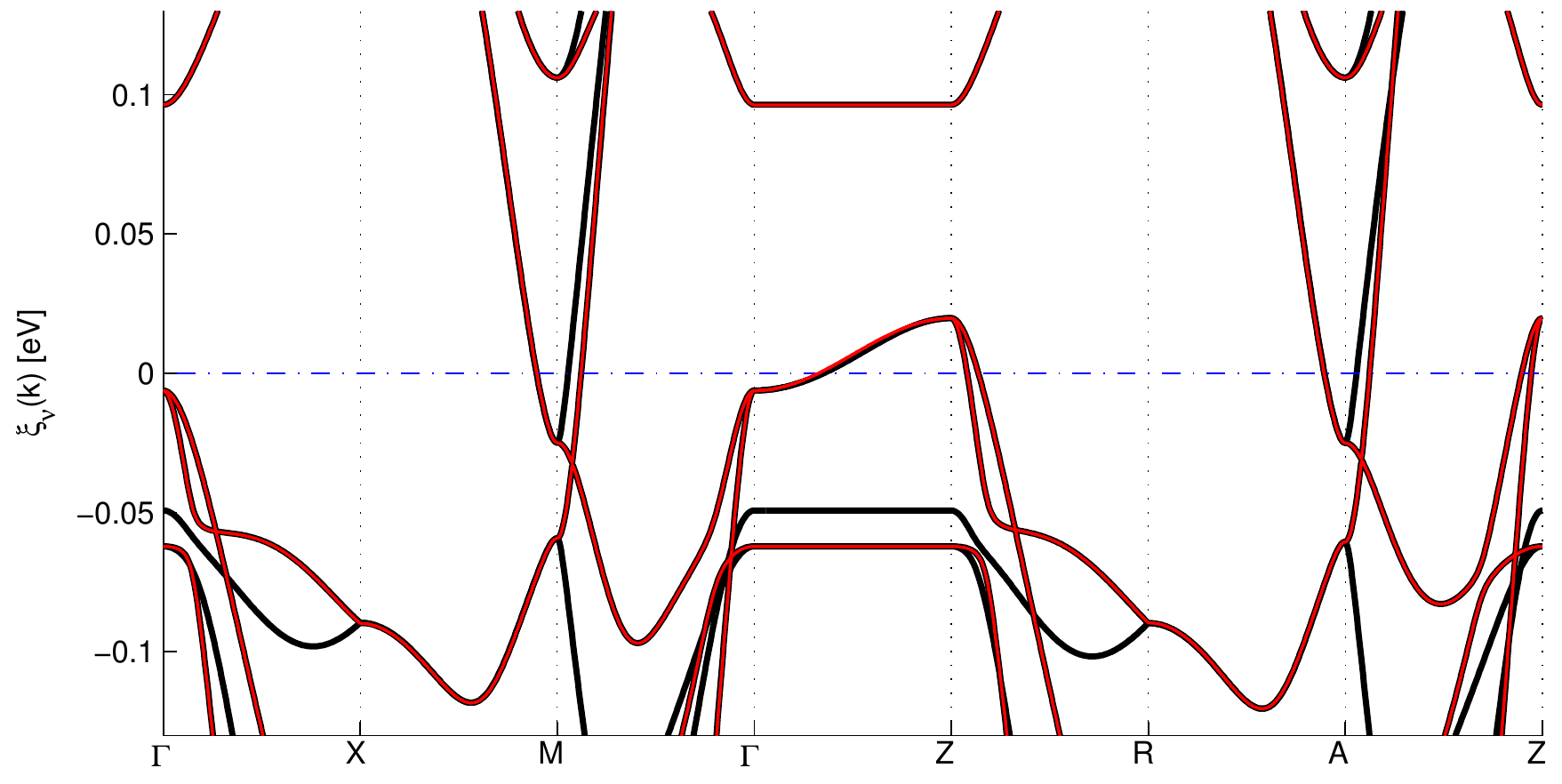}
\rput[tr](-0.95\columnwidth,0.48\columnwidth){(a)}\newline
\includegraphics[width=0.98\columnwidth]{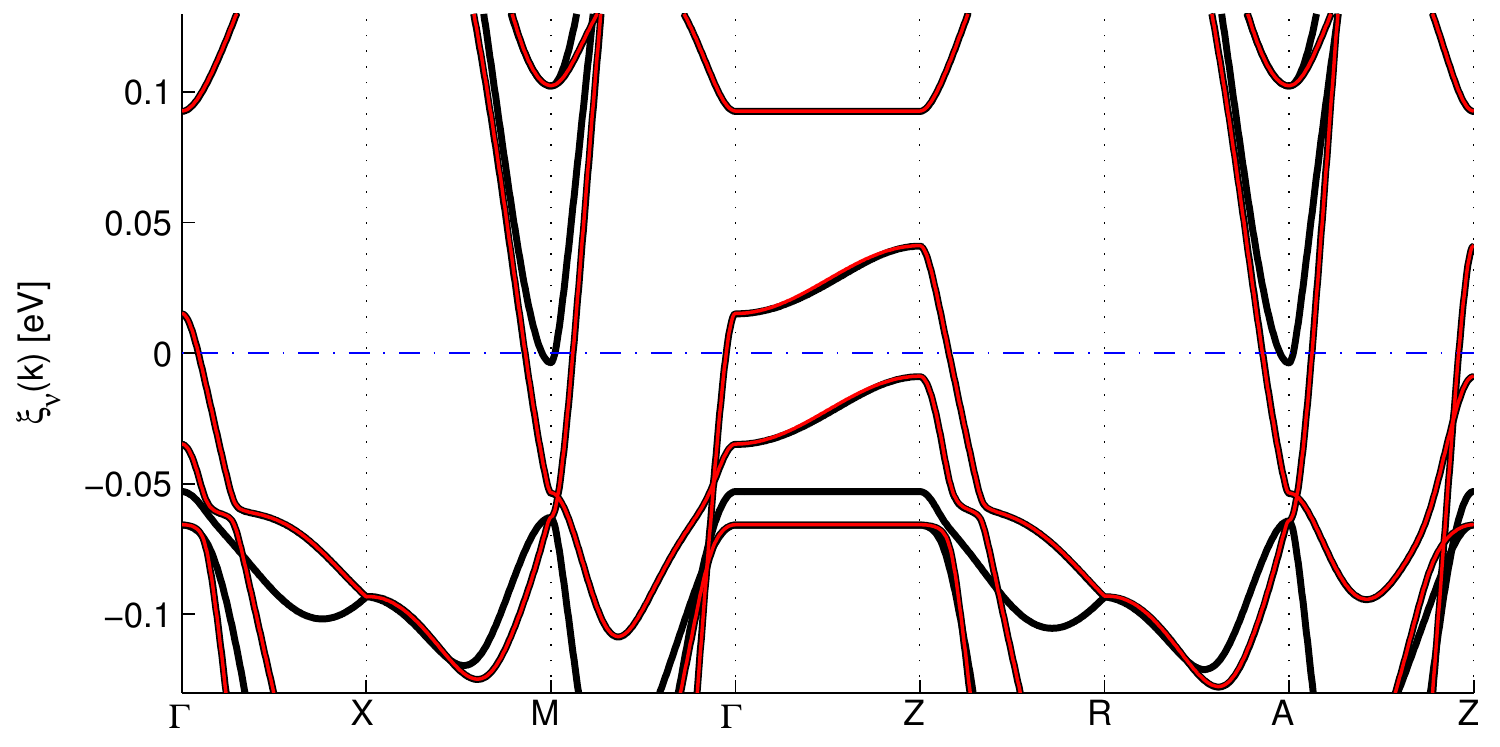}
\rput[tr](-0.95\columnwidth,0.48\columnwidth){(b)}\newline
\caption{(Color online) Comparison of the band structure from the tight-binding model from ten orbitals (black line) and
five orbitals (red line) (for $H_{\text{5Band}}=H^{++}+ H^{+-}$) showing the nearly perfect agreement in the high temperature tetragonal phase (a) and the low temperature orbitally ordered state (b).}
\label{fig:compare}
\end{figure}
\begin{gather*}
t^{11}_{11}=0.086/z, t^{10}_{16}=-0.063/z-0.0211,\\
t^{20}_{11}=-0.028/z+0.0028, t^{21}_{16}=0.017/z,\\
t^{11}_{13}=-0.056i/z, t^{10}_{18}=0.305i/z+0.076i,\\
t^{11}_{15}=-0.109/z, t^{10}_{27}=-0.412/z+0.022,\\
t^{11}_{22}=-0.066/z-0.003, t^{10}_{29}=-0.364i/z-0.034i,\\
t^{11}_{23}=0.089i/z, t^{10}_{2,10}=0.338/z-0.018,\\
t^{11}_{33}=0.232/z, t^{10}_{38}=0.080/z+0.0025,\\
t^{20}_{33}=0.009/z, t^{21}_{38}=0.016/z,\\
t^{02}_{33}=-0.045/z, t^{10}_{49}=0.311/z-0.002,\\
t^{22}_{33}=0.027/z, t^{21}_{49}=-0.019/z,\\
t^{11}_{34}=0.099/z, t^{10}_{4,10}=0.180i/z,\\
t^{11}_{35}=0.146i/z,\\
\epsilon_1=0.014/z,\\
\epsilon_2=-0.539/z+0.029,\\
\epsilon_3=0.020/z,\\
\epsilon_4=0.020/z,\\
\epsilon_5=-0.581/z+0.032
\end{gather*}
For the $k_z$ dispersion the additional hopping terms are,
\begin{gather*}
t^{101}_{16}=0+0.0027,\\
t^{001}_{11}=0, t^{211}_{16}=-0.017/z,\\
t^{111}_{11}=0, t^{101}_{18}=0.009i/z,\\
t^{201}_{11}=0.017/z-0.0027, t^{101}_{19}=0.020i/z,\\
t^{201}_{14}=0.0030i/z+0.0045i, t^{211}_{19}=-0.0031i/z,\\
t^{001}_{33}=0.011/z, t^{101}_{38}=0.006/z+0.002,\\
t^{201}_{33}=-0.008/z, t^{211}_{38}=-0.003/z,\\
t^{021}_{33}=0.020/z, t^{101}_{39}=0.015/z,\\
t^{011}_{49}=0.025/z-0.0020,\\
t^{121}_{49}=0.006/z.
\end{gather*}
The five orbital tight-binding model can be generated from the ten orbital parameters given above by taking the symmetric and antisymmetric combinations as given in Ref.\onlinecite{Eschrig09}. The five orbital model is given by
\begin{eqnarray}
 H_{\text{5Band}}=H^{++}\pm H^{+-}.
\end{eqnarray}
The tight-binding parameters of the five orbital 3D model for the real orbitals ($xy$,$x^2-y^2$,$xz$,$yz$,$z^2$) are given by
\begin{gather*}
t^{11}_{11}=0.086/z, t^{10}_{16}=-0.063/z-0.0211,\\
t^{20}_{11}=-0.028/z+0.0028, t^{21}_{16}=0.017/z,\\
t^{11}_{13}=-0.056i/z, t^{10}_{18}=0.305i/z+0.076i,\\
t^{11}_{15}=-0.109/z, t^{10}_{27}=-0.412/z+0.022,\\
t^{11}_{22}=-0.066/z-0.003, t^{10}_{29}=-0.364i/z-0.034i,\\
t^{11}_{23}=0.089i/z, t^{10}_{2,10}=0.338/z-0.018,\\
t^{11}_{33}=0.232/z, t^{10}_{38}=0.080/z+0.0025,\\
t^{20}_{33}=0.009/z, t^{21}_{38}=0.016/z,\\
t^{02}_{33}=-0.045/z, t^{10}_{49}=0.311/z-0.002,\\
t^{22}_{33}=0.027/z, t^{21}_{49}=-0.019/z,\\
t^{11}_{34}=0.099/z, t^{10}_{4,10}=-0.180i/z,\\
t^{11}_{35}=-0.146i/z,\\
\epsilon_1=0.014/z,\\
\epsilon_2=-0.539/z+0.029,\\
\epsilon_3=0.020/z,\\
\epsilon_4=0.020/z,\\
\epsilon_5=-0.581/z+0.032.
\end{gather*}
For the $k_z$ dispersion the additional hopping terms are
\begin{gather*}
t^{101}_{16}=0.0+0.0165,\\
t^{001}_{11}=0.0, t^{211}_{16}=0.0,\\
t^{111}_{11}=0.0, t^{101}_{18}=0.0,\\
t^{201}_{11}=0.017/z-0.0165, t^{101}_{19}=0.020/z,\\
t^{201}_{14}=0.0030/z+0.027, t^{211}_{19}=0.0,\\
t^{001}_{33}=0.011/z, t^{101}_{38}=0.006/z+0.0152,\\
t^{201}_{33}=-0.008/z, t^{211}_{38}=0.0,\\
t^{021}_{33}=0.0, t^{101}_{39}=0.015/z,\\
t^{011}_{49}=0.025/z-0.0122,\\
t^{121}_{49}=0.0.
\end{gather*}
The above five orbital fit agrees well with the ten orbital model as can be seen from Fig.\ref{fig:compare}.
In Fig.\ref{fig:QOSO} we show the QO results for magnetic fields along two orthogonal in-plane directions. As can be seen in Fig.\ref{fig:QOSO} the $k_z$ dispersion ($\theta$ dependence) of the orbit areas are slightly different for $\phi=0^{\circ}$ and $\phi=90^{\circ}$. These two $\phi$ values would correspond to the areas observed from the two structural domains for a given magnetic field direction in quantum oscillation experiment and therefore have slightly different QO frequencies when $\theta \neq 0$.
\begin{figure}[tb]
\includegraphics[width=0.98\columnwidth]{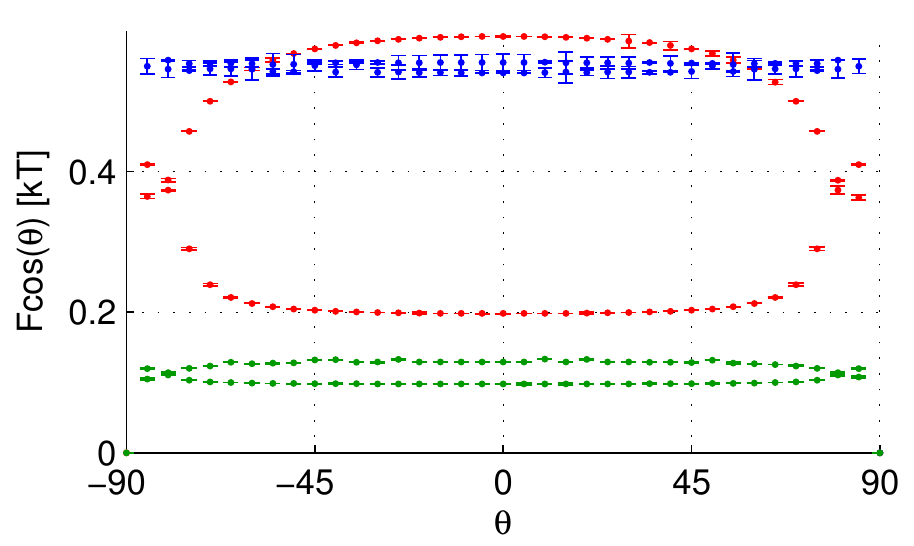}
\rput[tr](-0.245\columnwidth,0.41\columnwidth){\includegraphics[width=0.4\columnwidth]{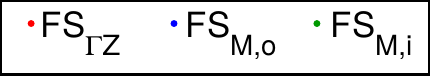}}
\rput[tr](-0.94\columnwidth,0.55\columnwidth){(a)}\newline
\includegraphics[width=0.98\columnwidth]{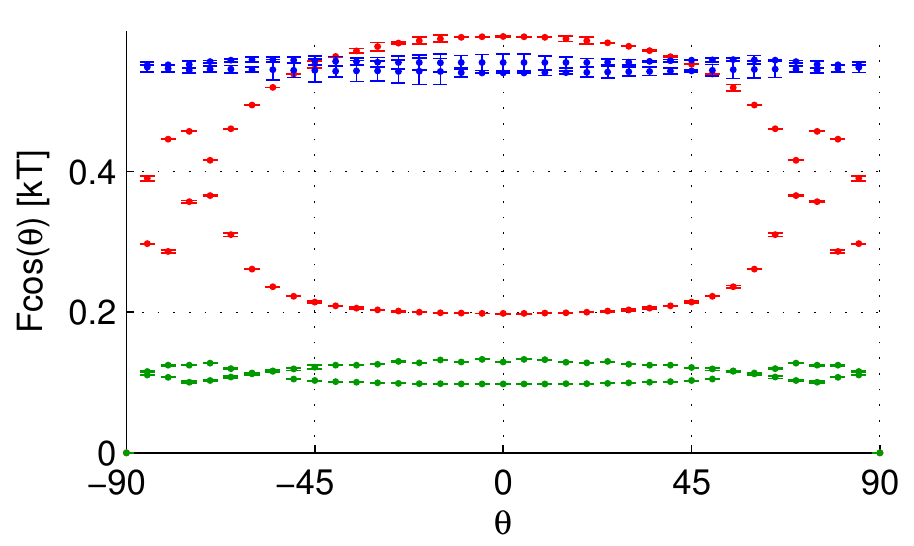}
\rput[tr](-0.94\columnwidth,0.55\columnwidth){(b)}\newline
\caption{(Color online) Extremal orbit area of Fermi pocket at low temperatures in the absence of SO coupling. The plots are for various $\theta$ values and shows the small difference in the k$_z$ dispersion of the Fermi pockets between $\phi=0^{\circ}$ (a) and $\phi=90^{\circ}$ (b). Here $\theta$ is the angle with c-axis and $\phi=0^{\circ}$ is aligned along the Fe-Fe direction.}
\label{fig:QOSO}
\end{figure}
\begin{figure}[tb]
\includegraphics[width=0.98\columnwidth]{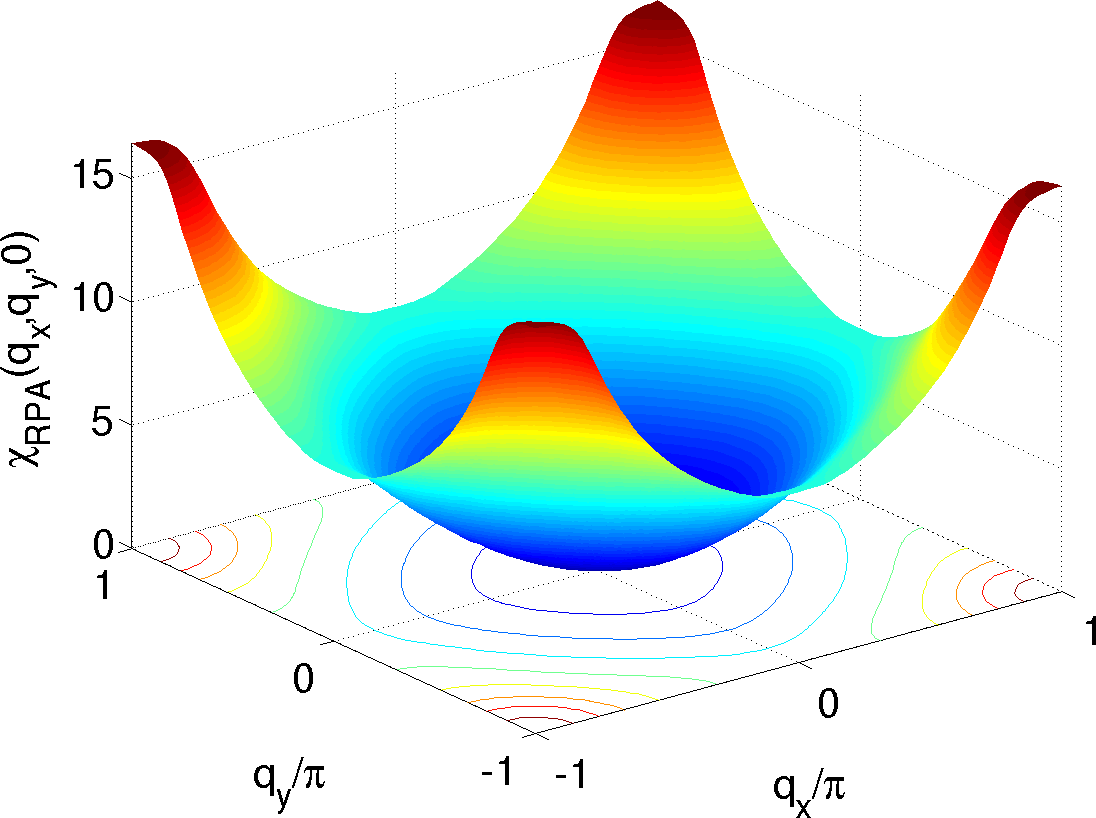}
\rput[tr](-0.95\columnwidth,0.65\columnwidth){(a)}\newline
\includegraphics[width=0.98\columnwidth]{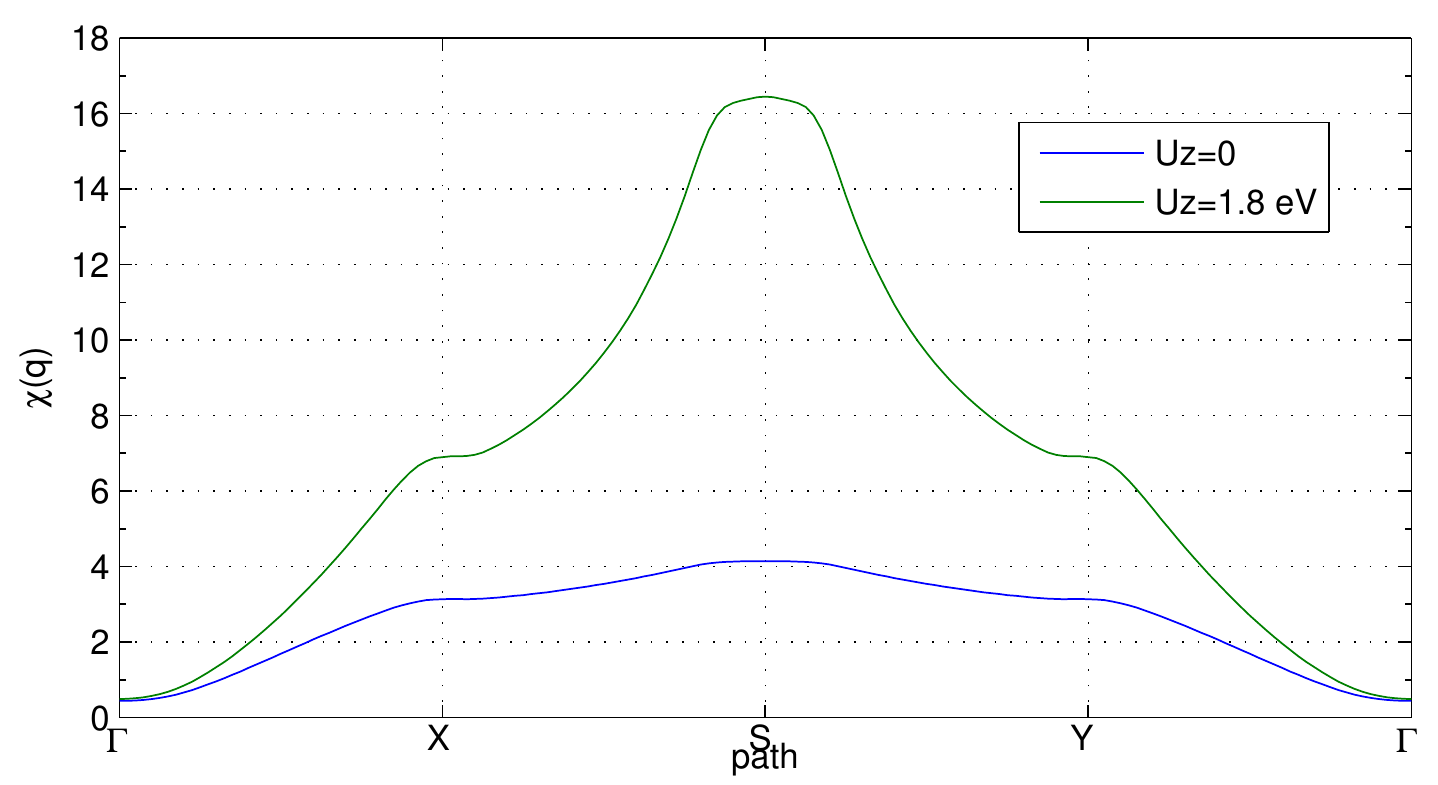}
\rput[tr](-0.95\columnwidth,0.5\columnwidth){(b)}
\caption{(Color online) RPA susceptibility in the tetragonal phase at $T=110\,\text{K}$ calculated for $Uz=1.8\,\text{eV}$ and $Jz=0.1Uz$ and plotted in the Brillouin zone at $k_z=0$ (a) and along a high symmetry path in comparison with the non-interacting susceptibility (b).}
\label{fig:RPA_T110}
\end{figure}
\begin{figure}[tb]
\includegraphics[width=0.98\columnwidth]{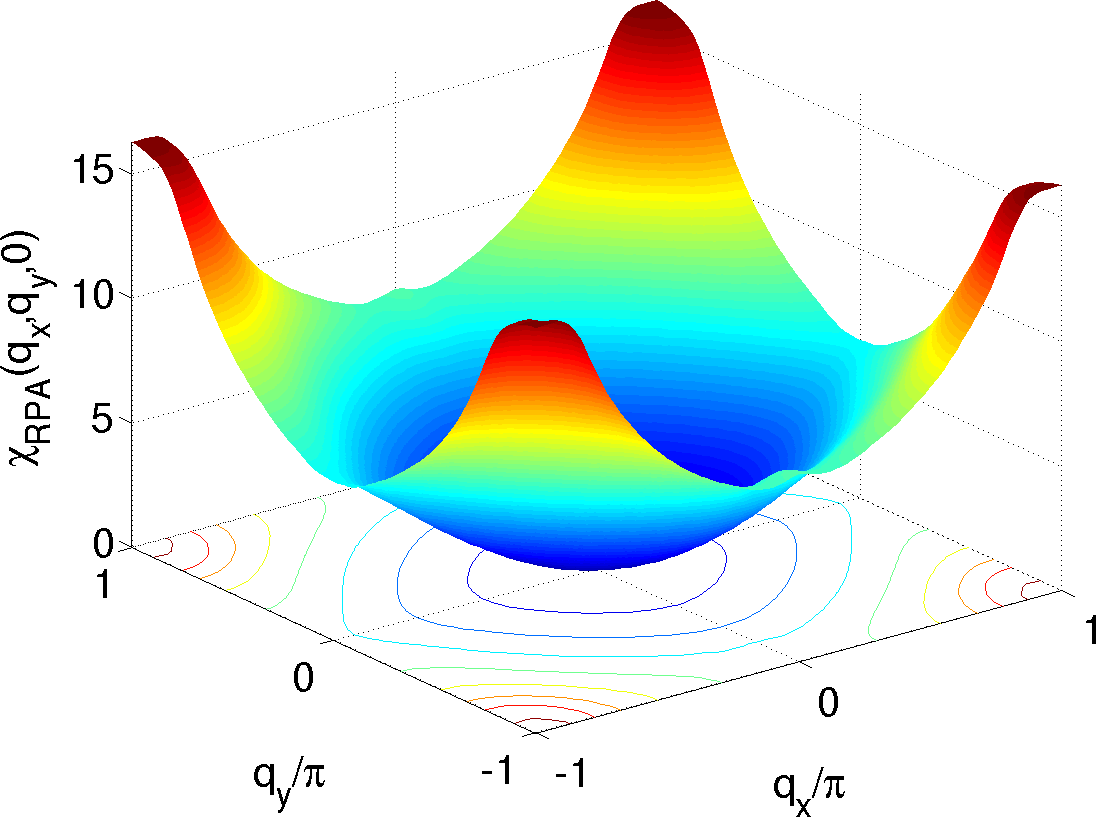}
\rput[tr](-0.95\columnwidth,0.65\columnwidth){(a)}\newline
\includegraphics[width=0.98\columnwidth]{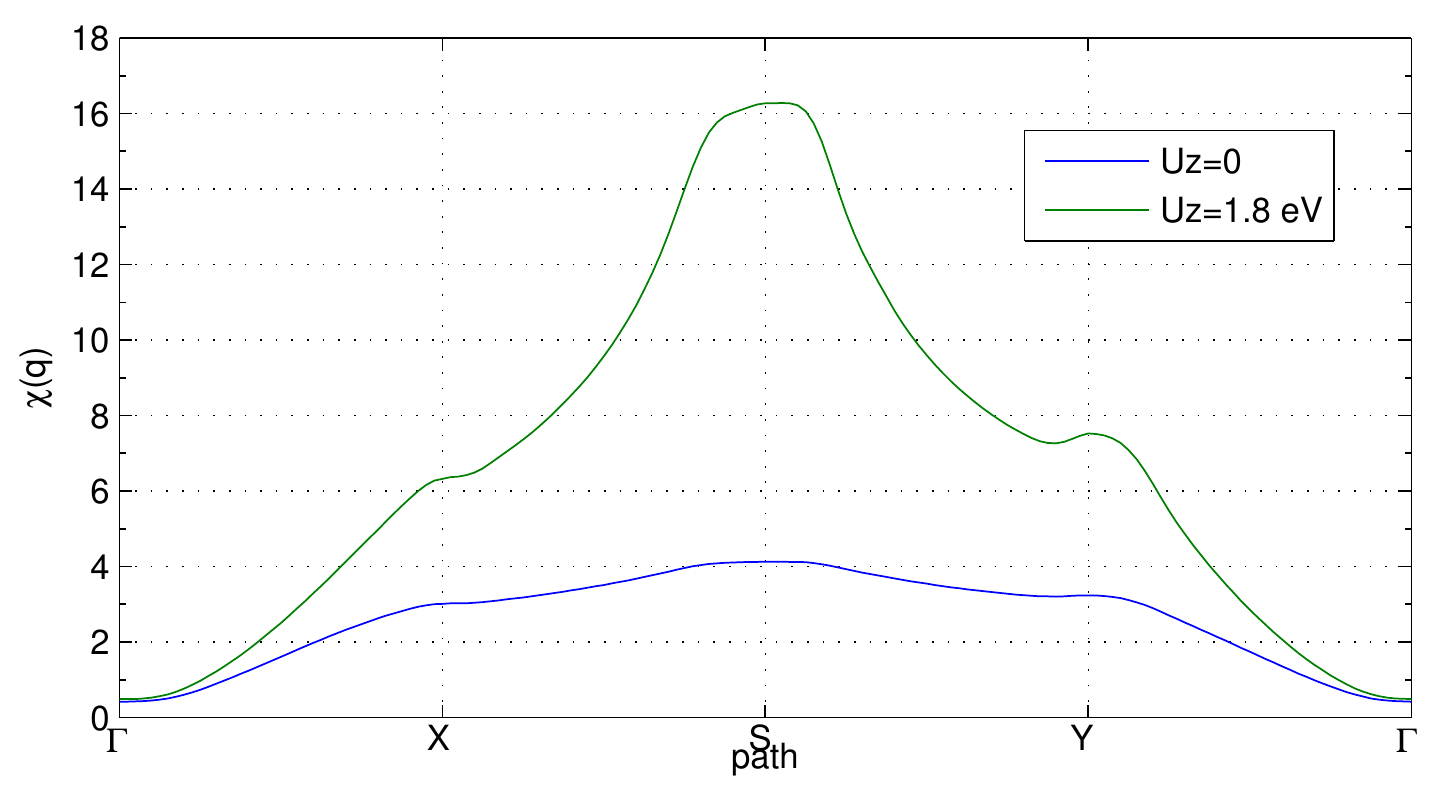}
\rput[tr](-0.95\columnwidth,0.5\columnwidth){(b)}
\caption{(Color online) RPA susceptibility in the orbital ordered phase at $T=40\,\text{K}$ (b) calculated for $Uz=1.8\,\text{eV}$ and $Jz=0.1Uz$ and plotted in the Brillouin zone at $k_z=0$ (a) and along a high symmetry path in comparison with the non-interacting susceptibility (b).}
\label{fig:RPA_T40}
\end{figure}

\textit{Interaction Hamiltonian.}
The local interactions are included via the Hubbard-Hund Hamiltonian
\begin{eqnarray}
	H = H_{0}& + &{U}\sum_{i,\ell}n_{i\ell\uparrow}n_{i\ell\downarrow}+{U}'\sum_{i,\ell'<\ell}n_{i\ell}n_{i\ell'}
	\nonumber\\
	& + & {J}\sum_{i,\ell'<\ell}\sum_{\sigma,\sigma'}c_{i\ell\sigma}^{\dagger}c_{i\ell'\sigma'}^{\dagger}c_{i\ell\sigma'}c_{i\ell'\sigma}\\
	& + & {J}'\sum_{i,\ell'\neq\ell}c_{i\ell\uparrow}^{\dagger}c_{i\ell\downarrow}^{\dagger}c_{i\ell'\downarrow}c_{i\ell'\uparrow} \nonumber \label{H},
\end{eqnarray}
where the interaction parameters ${U}$, ${U}'$, ${J}$, ${J}'$ are given in the notation of Kuroki \textit{et al.} \cite{Kuroki08}. Here, $\ell$ is an orbital index with $\ell\in(1,\ldots,5)$ corresponding to the Fe-orbitals. The pairing interaction in band representation is now calculated from the interaction in the orbital representation via
\begin{eqnarray}
	{\Gamma}(\k,\k') & = & \mathrm{Re}\sum_{\ell_1\ell_2\ell_3\ell_4} a_{\nu_i}^{\ell_1,*}(\k) a_{\nu_i}^{\ell_4,*}(-\k) \\
	&&\times \left[{\Gamma}_{\ell_1\ell_2\ell_3\ell_4} (\k,\k') \right] a_{\nu_j}^{\ell_2}(\k') a_{\nu_j}^{\ell_3}(-\k')\,,\nonumber \label{eq:Gam_ij}
\end{eqnarray}
where $\k$ and $\k'$ are quasiparticle momenta restricted to the pockets $\k \in \text{FS}_i$ and $\k' \in \text{FS}_j$, where $i$ and $j$ correspond to the band index of the  Fermi surface sheets. The vertex function in orbital space $\Gamma_{\ell_1\ell_2\ell_3\ell_4}$ describes the particle-particle scattering of electrons in orbitals $\ell_2,\ell_3$ into $\ell_1,\ell_4$.
\begin{figure}[tb]
 \flushleft\includegraphics[width=0.5\columnwidth]{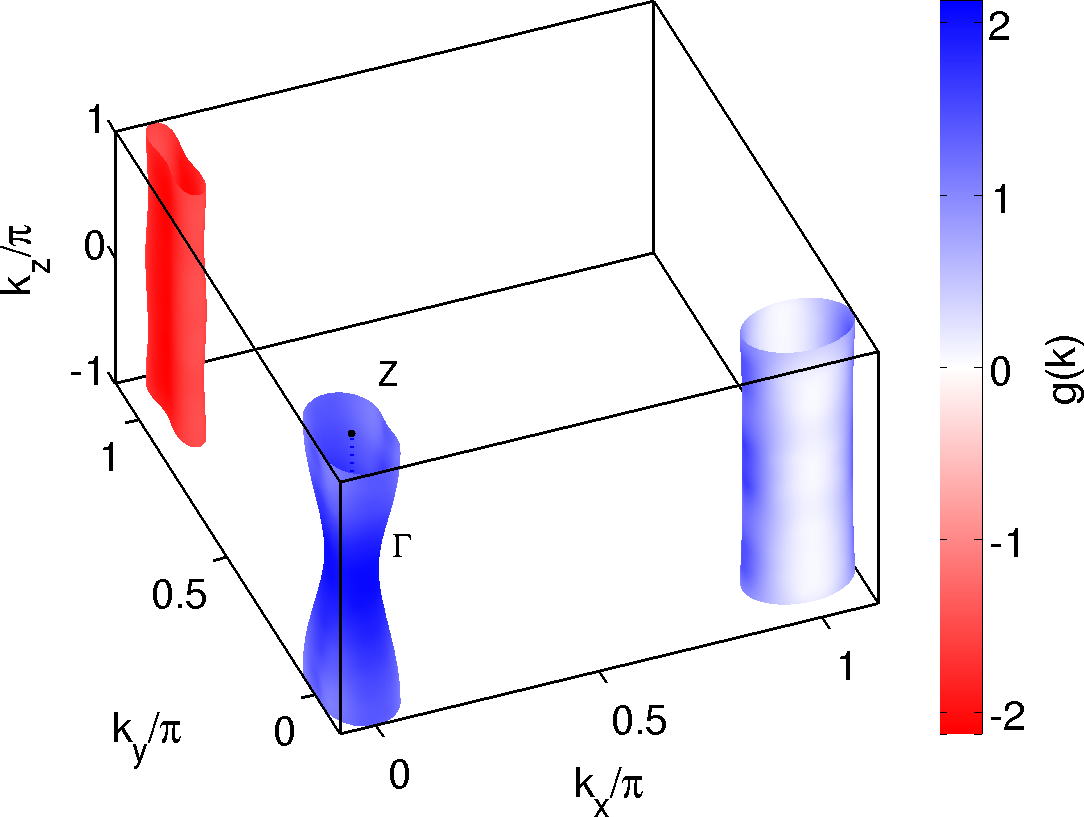}
  \rput[tr](-0.45\columnwidth,0.4\columnwidth){(a)}
\rput[tr](0.52\columnwidth,0.42\columnwidth){\includegraphics[width=0.52\columnwidth]{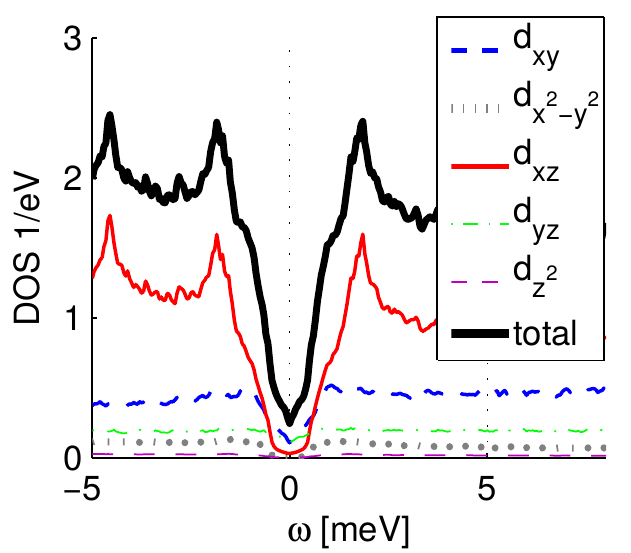}}
\rput[tr](0.03\columnwidth,0.4\columnwidth){(b)}
\caption{(Color online) Superconducting gap function plotted over the Fermi surface of the 5 orbital model for $Uz=1.92\,\text{eV}$ and $Jz=0.125Uz$ (a) and corresponding density of states (b).}
\label{fig:Gap}
\end{figure}

In RPA it is given by
\begin{align}
	&{\Gamma}_{\ell_1\ell_2\ell_3\ell_4} (\k,\k') = \left[\frac{3}{2} \bar U^s \chi_1^{\rm RPA} (\k-\k') \bar U^s\nonumber \right.\,~~~~~~\,\\
	&\quad\left. +  \frac{1}{2} \bar U^s - \frac{1}{2}\bar U^c \chi_0^{\rm RPA} (\k-\k') \bar U^c + \frac{1}{2} \bar U^c \right]_{\ell_1\ell_2\ell_3\ell_4}. \label{eq:fullGamma}
\end{align}
Note further, that in the spin-singlet channel, we symmetrize the pairing vertex by using the expression $1/2[ \Gamma(\k,\k')+\Gamma(\k,-\k')]$ in the linearized gap equation\cite{s_graser_09}. The RPA susceptibility in ${\bf{q}}$-space is shown in Figs.\ref{fig:RPA_T110} and \ref{fig:RPA_T40}. The $\omega=0$ susceptibility is dominated by the $(\pi,\pi)$ nesting between the electron Fermi pockets.  Note, however, that due to the small band energies for this system, the 
 ${\bf q}$-dependence at relatively low finite $\omega$ can be quite different, as we will explore elsewhere. The solution of the linearized gap equation for $Uz=1.92\,\text{eV}$ and $Jz=0.125Uz$ is plotted in Fig.\ref{fig:Gap}. It can be seen that in comparison to the gap structure obtained for $Uz=1.8\,\text{eV}$ and $Jz=0.1Uz$ in the manuscript, the gap structure for a larger $U$ leads to a weakening of the nodes on the electron pocket. The densities of states $\rho(\omega)$ are then calculated using the gap $\Delta(\k)=\Delta_0 g(\k)$ interpolated on a fine $k$-mesh of approximately 1000$\times$1000$\times$200 points and an imaginary smearing of $\eta=0.02 \,\text{meV}$.

Finally, the effect of the density of states on the penetration depth is $\lambda$ is estimated by the formula
\begin{align}
 \lambda\propto \int d\omega \rho(\omega) \frac{df(\omega)}{d\omega},
\end{align}
where $f(\omega)$ is the Fermi function.

\begin{figure}[tb]
\includegraphics[width=0.98\columnwidth]{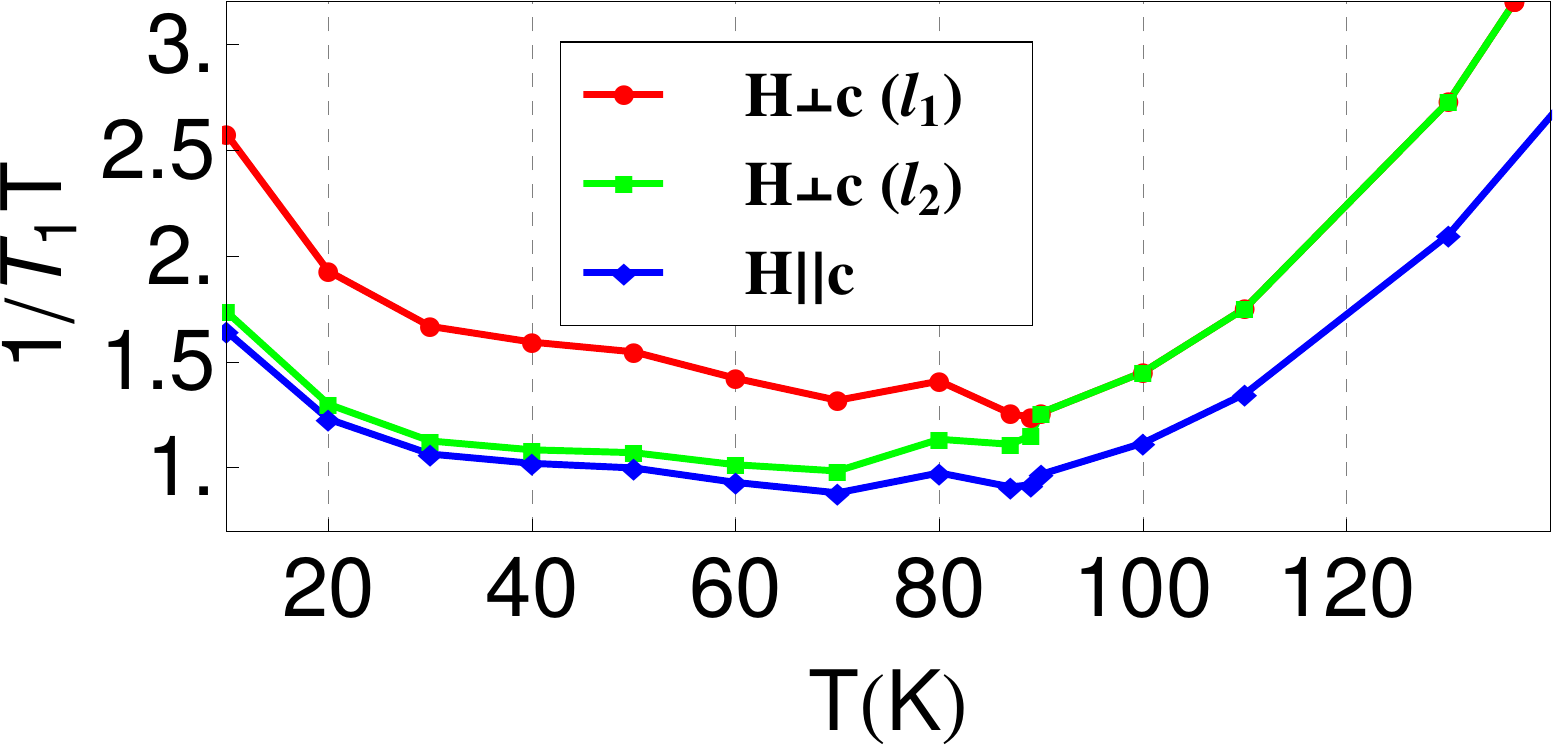}
 \caption{(Color online) Spin lattice relaxation rate calculated in the absence of structure factor filtering i.e A$_{hf}({\bf{q}})=1$ for the same parameters as used in Fig.3 of the manuscript.}
\label{fig:t1tnosf}
\end{figure}

\textit{Spin Lattice Relaxation Rate.}
The NMR spin lattice relaxation rate is given as
\begin{eqnarray}
\frac{1}{T_1T}\propto \lim_{\omega\rightarrow 0}\sum_{\bf{q}}|A_{hf}({\bf{q}})|^2\frac{\mathop{\text{Im}}[\chi_{RPA}(\bf{q},\omega)]}{\omega},
\end{eqnarray}
where $\chi_{RPA}({\bf{q}},\omega)$ is the RPA enhanced susceptibility and $A_{hf}({\bf{q}})=A_{hf}\cos(q_x)\cos(q_y)$ is the hyperfine form factor which is a $3\times3$ matrix with non zero diagonal elements in the paramagnetic state. The ${\bf{q}}$ dependence in the form factor leads to filtering of the fluctuations at the Brillouin zone edges. Comparing the relaxation rate given in Fig.3(b) in the manuscript with the corresponding case without the form factor (see Fig.\ref{fig:t1tnosf}) we find that in the absence of filtering for $A_{hf}({\bf{q}})=1$, the upturn in ${1/T_1T}$ takes place at a higher temperature.

For the multi-orbital model, the expression for $1/T_1T$ can be simplified by expanding the bare susceptibility in powers of $\omega$. After some algebra, this leads to the following expression,
\begin{figure}[tb]
\includegraphics[width=0.977\columnwidth]{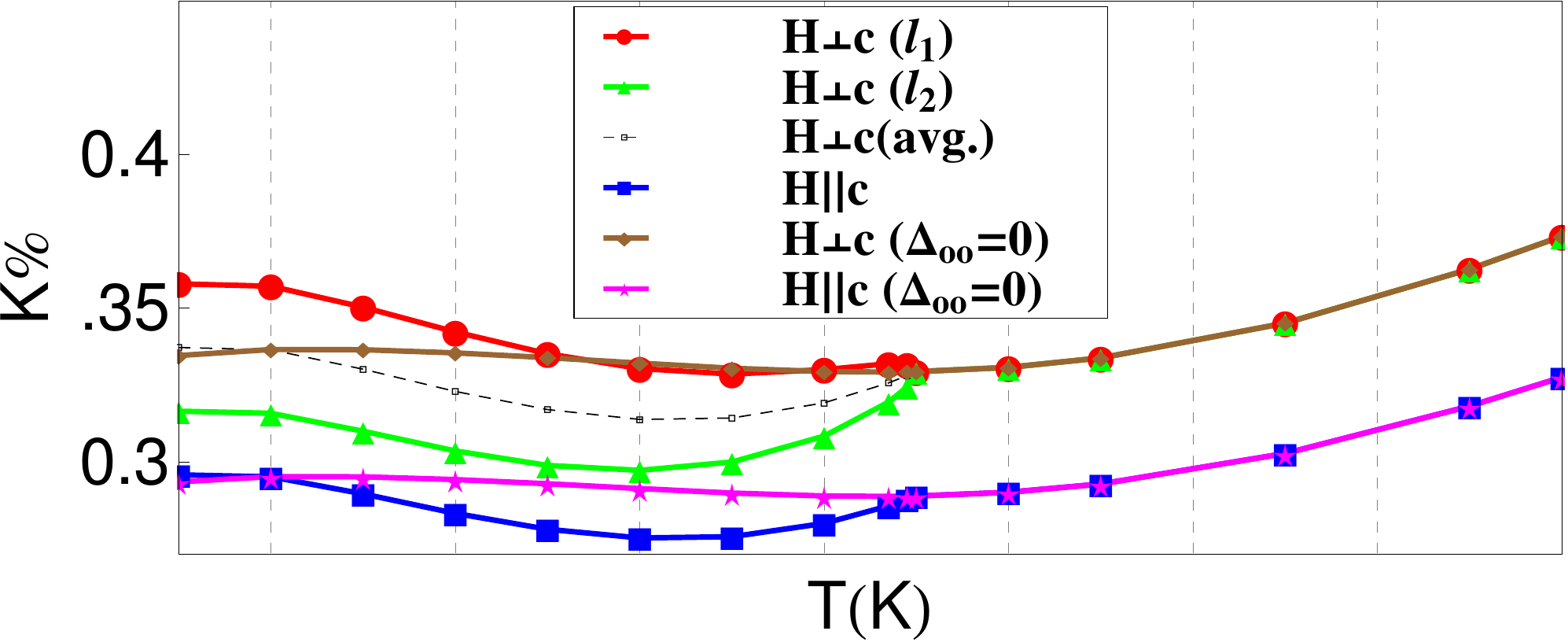}
\rput[tr](-0.95\columnwidth,0.4\columnwidth){(a)}\newline
\includegraphics[width=0.977\columnwidth]{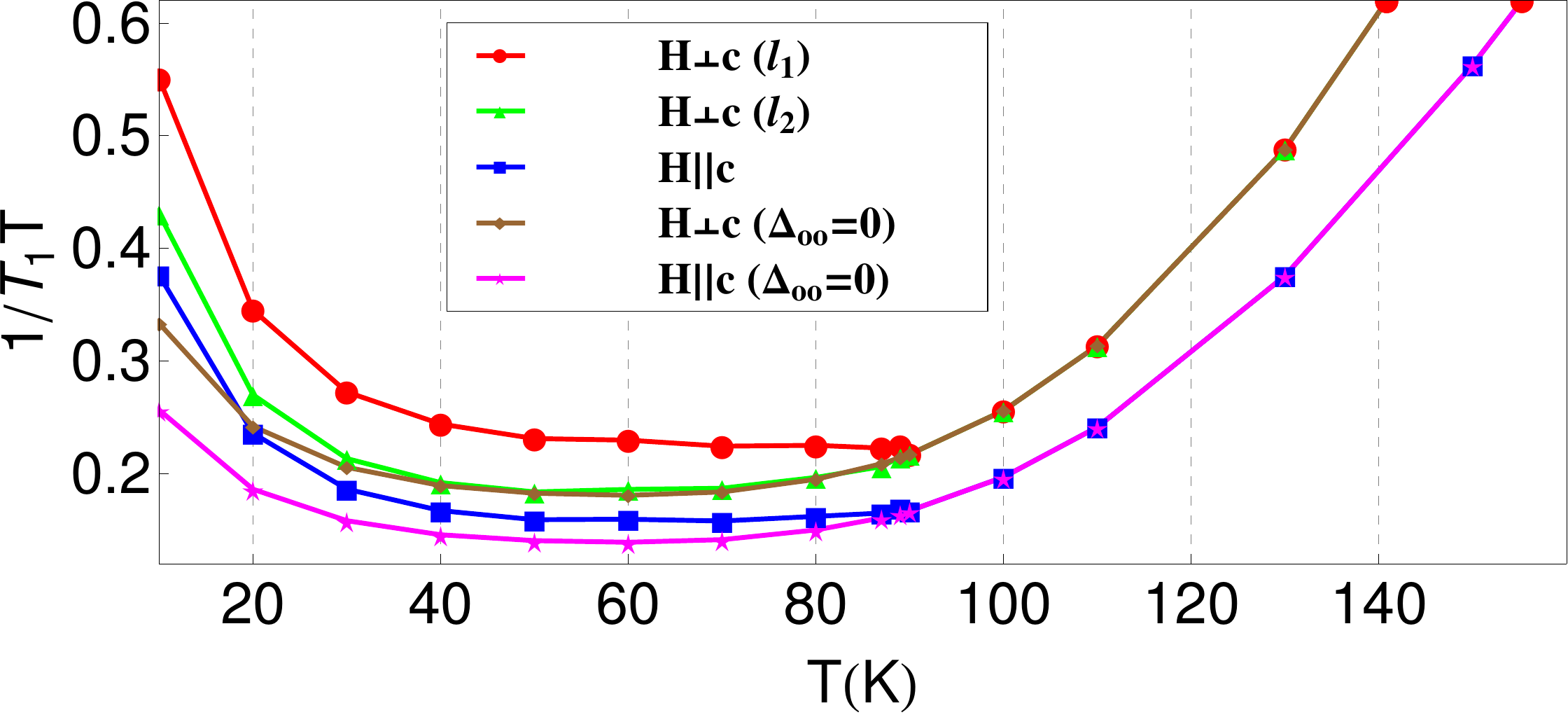}
\rput[tr](-0.95\columnwidth,0.45\columnwidth){(b)}\newline
\caption{(Color online) Plot showing the effects of orbital ordering on spin-lattice relaxation rate and NMR Knight shift. The hyperfine form factor $A_{hf}=\alpha+\beta (T)$ are same as in manuscript Fig.3 except the temperature dependent factor $\beta (T)=0$ in the absence of orbital order. (a) NMR Knight shift versus $T$ comparing cases with and without orbital order. (b) Spin-lattice relaxation rate versus $T$ for cases with and without orbital order. }
\label{fig:knightcompare}
\end{figure}
\begin{eqnarray}
\frac{1}{T_1T}=\sum_{\text{mat.el.}}\sum_{{\bf{q}}}|A_{hf}({\bf{q}})|^2{\mathop{\text{Re}}}
\left(\frac {\hat{B}({\bf{q}})} {\hat{1}-\hat{U}\hat{A}({\bf{q}})}\nonumber \right. \\
\left. +\frac{\hat{A}({\bf{q}})}{\hat{1}-\hat{U}\hat{A}({\bf{q}})}\hat{U}\hat{B}({\bf{q}})\frac{\hat{1}}{\hat{1}-\hat{U}\hat{A}({\bf{q}})}\right),
\end{eqnarray}
where the matrices $\hat{A}$, and $\hat{B}$ are given by
\begin{eqnarray}
\hat{A}({\bf{q}})&=&-\frac{1}{2N}\sum_{{\bf{k}},\mu\nu}\hat{M}_{\mu,\nu}({\bf{k}},{\bf{q}}){\mathop{\text{Re}}}(G_{\mu,\nu}({\bf{k}},{\bf{q}},0))\nonumber\\
&&\times [f(E_{\nu}({\bf{k}}+{\bf{q}}))-f(E_{\mu}({\bf{k}}))]\,,\\
\hat{B}({\bf{q}})&=&-\frac{1}{2N}\sum_{{\bf{k}},\mu\nu}\hat{M}_{\mu,\nu}({\bf{k}},{\bf{q}}){\mathop{\text{Im}}}\left(\frac{\partial G_{\mu,\nu}({\bf{k}},{\bf{q}},\omega)}{\partial \omega}\right)_{\omega=0}\nonumber\\
&&\times[f(E_{\nu}({\bf{k}}+{\bf{q}}))-f(E_{\mu}({\bf{k}}))]\,.
\end{eqnarray}
Here $\hat{M}_{\mu,\nu}({\bf{k}},{\bf{q}})=a_{\mu}^s({\bf{k}})a_{\mu}^{p*}({\bf{k}})a_{\nu}^q({{\bf{k}}+\bf{q}})a_{\nu}^{t*}({\bf{k}}+{\bf{q}})$ is in general complex with $a_{\mu}^s({\bf{k}})$ representing the eigenvector component corresponding to band $\mu$ and orbital $s$, $E_{\mu}$ is the eigenvalue for band $\mu$, and $f(E_{\mu}({\bf{k}}))$ is the Fermi function at energy $E_{\mu}({\bf{k}})$. The Greens function reads
\begin{align}
G_{\mu,\nu}({\bf{k}},{\bf{q}},\omega)=\frac{1}{E_{\nu}({\bf{k}}+{\bf{q}})-E_{\mu}({\bf{k}})-\omega-i\delta}.
\end{align}
The above expressions for Knight shift and spin lattice relaxation rate have been evaluated for our band structure in Fig.3 of the manuscript. The effect of orbital ordering can be made clearer upon comparing the cases with and without orbital ordering. As shown in Fig.\ref{fig:knightcompare} (a), we find that in the absence of orbital ordering the NMR Knight shift saturates as we go to low temperatures. The spin lattice relaxation rate in Fig.\ref{fig:knightcompare} (b) shows that although band structure effects do cause a weak enhancement of spin fluctuations even in the absence of orbital ordering, the effect is enhanced due to the presence of orbital ordering.

\textit{s+d wave orbital order.}

The effect of an additional $d$-wave orbital order can be included with the term
\begin{eqnarray}
\!\!\!&&H_{OO}=\sum_{{\bf{k}}\sigma}[\Delta_{s}(T)(n_{xz\sigma}({\bf{k}})-n_{yz\sigma}({\bf{k}}))+\nonumber\\
\!\!\!&&\Delta_{d}(T)(\cos(k_x)-\cos(k_y))(n_{xz\sigma}({\bf{k}})+n_{yz\sigma}({\bf{k}}))],
\end{eqnarray}
where both $\Delta_{s}(T)$ and $\Delta_{d}(T)$ exhibit a typical mean-field $T$ dependence and have been chosen such that $\Delta_s/\Delta_d>0$. The band structure shifts presented above have to be slightly adjusted to agree with this orbital order. The additional band shifts are
\begin{gather*}
\Delta t^{101}_{33}=\Delta t^{011}_{33}=0.00016,\\
\Delta t^{10}_{33}=\Delta t^{01}_{33}=-0.0013,\\
\Delta t^{10}_{11}=-\Delta t^{20}_{11}=-0.0006.
\end{gather*}

\begin{figure}[t]
\includegraphics[width=\columnwidth]{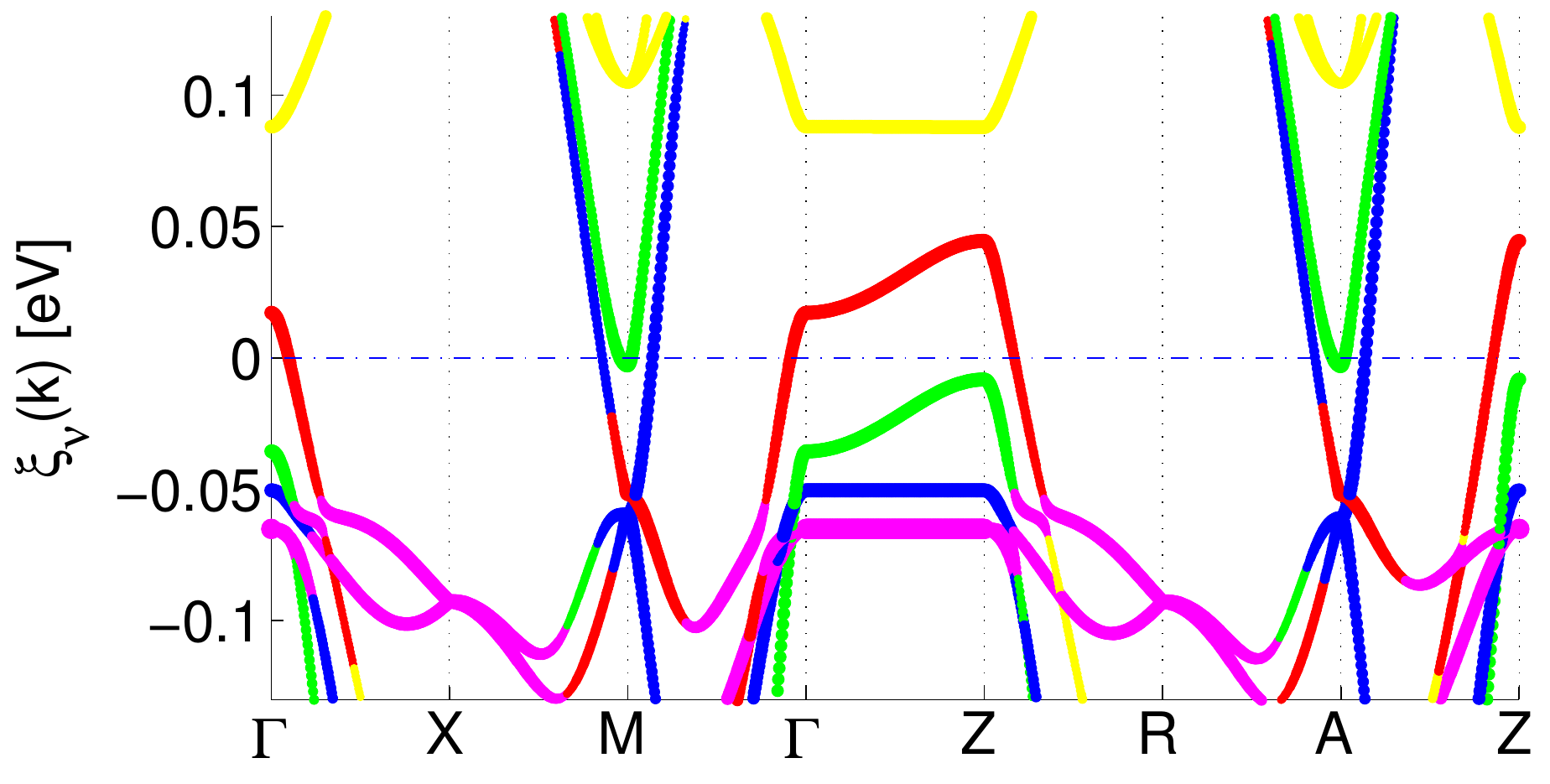}
\rput[tr](-0.3\columnwidth,0.55\columnwidth){(a)}\newline
\includegraphics[width=\columnwidth]{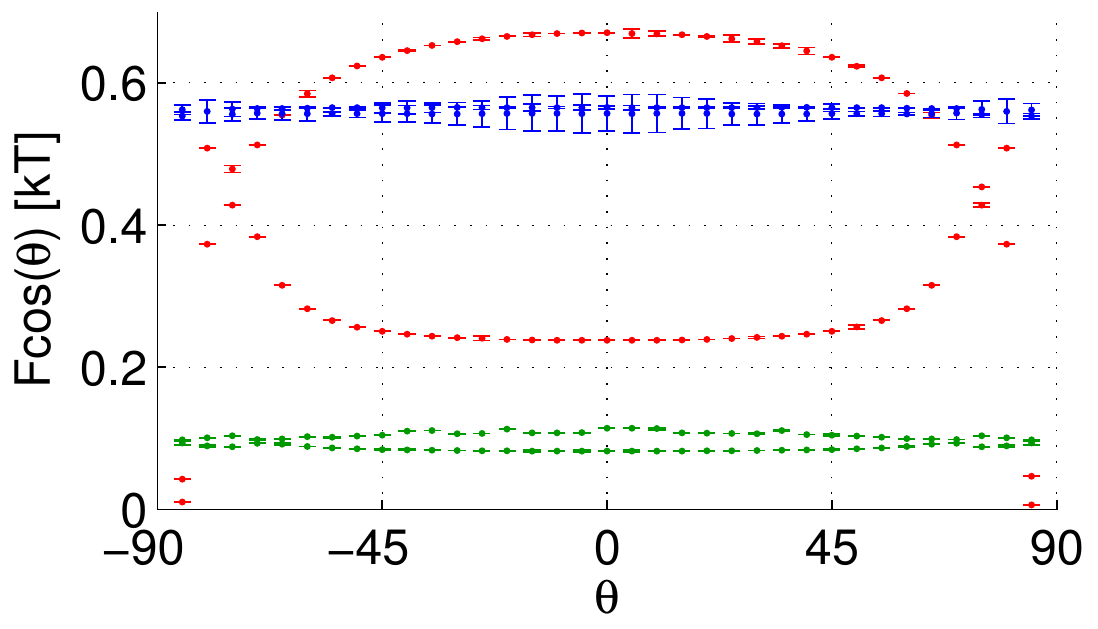}
\rput[tr](-0.45\columnwidth,0.6\columnwidth){(b)}
\caption{(Color online) (a) Band structure of the 10 orbital model at low $T$ with $d$-wave orbital order and SO coupling, yielding the QO frequencies as a function of magnetic field angle $\theta$ shown in (b), where the error bars indicate the numerical uncertainty in the determination of the extremal orbits. The orbital character in the band plot is indicated by the colors red $d_{xz}$, green $d_{yz}$, blue $d_{xy}$, yellow $d_{x^2-y^2}$, purple  $d_{3z^2-r^2}$.
}
\label{fig:QOdwave}
\end{figure}
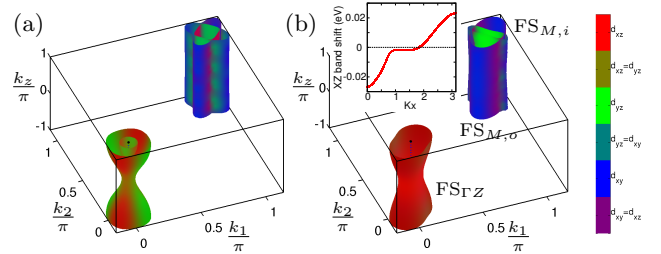
\begin{figure}[b]
\quad
\includegraphics[width=0.38\columnwidth]{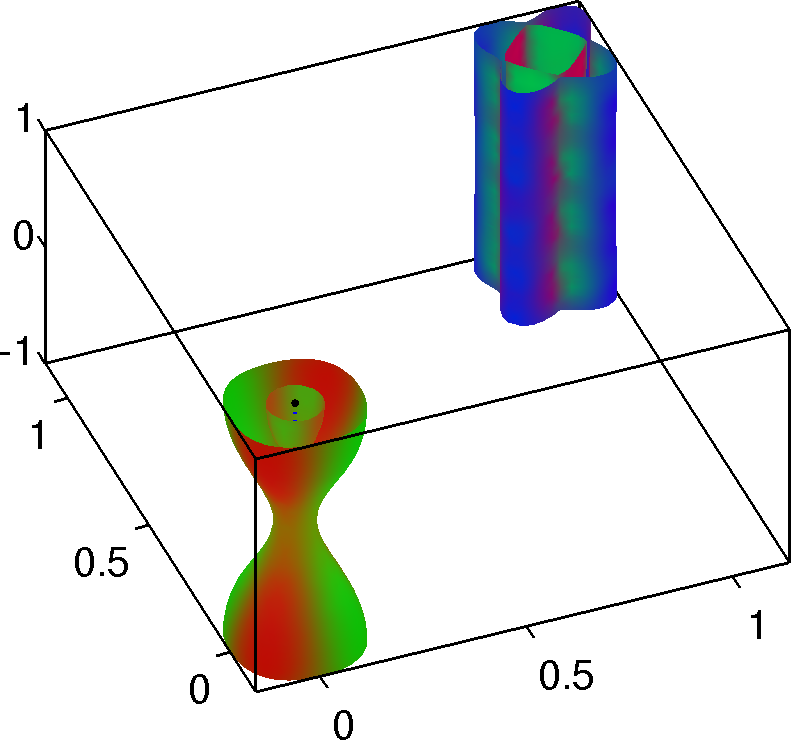}
\rput[tr](-0.38\columnwidth,0.36\columnwidth){(a)}
\rput[tr](-0.39\columnwidth,0.27\columnwidth){$\frac{k_z}\pi$}
\rput[tr](-0.33\columnwidth,0.07\columnwidth){$\frac{k_2}\pi$}
\rput[tr](-0.06\columnwidth,0.04\columnwidth){$\frac{k_1}\pi$}
\quad
\includegraphics[width=0.5\columnwidth]{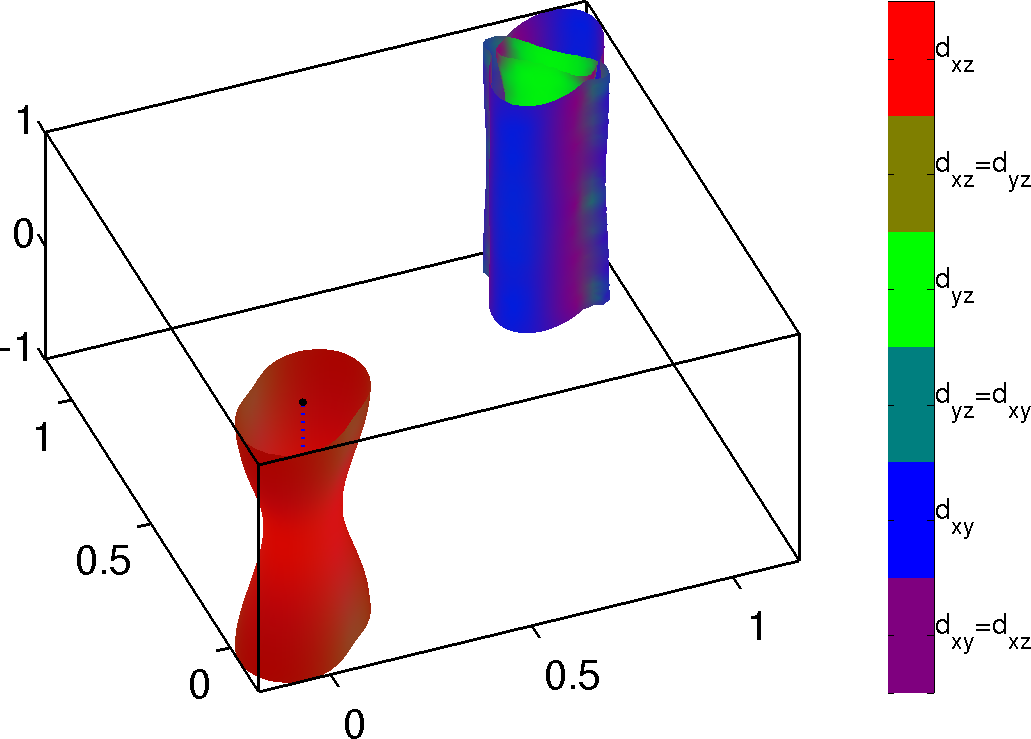}
\rput[tr](-0.29\columnwidth,0.38\columnwidth){\includegraphics[width=0.2\columnwidth]{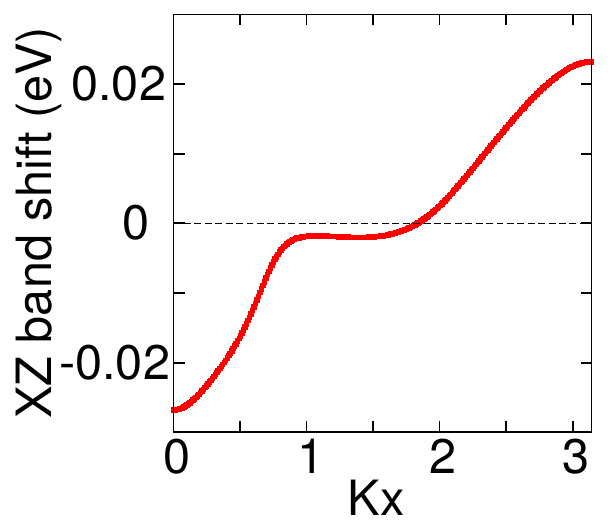}}
\rput[tr](-0.51\columnwidth,0.36\columnwidth){(b)}
\rput[tr](-0.51\columnwidth,0.27\columnwidth){$\frac{k_z}\pi$}
\rput[tr](-0.45\columnwidth,0.07\columnwidth){$\frac{k_2}\pi$}
\rput[tr](-0.18\columnwidth,0.04\columnwidth){$\frac{k_1}\pi$}
\rput[tr](-0.120\columnwidth,0.35\columnwidth){\scriptsize {FS$_{M,i}$}}
\rput[tr](-0.20\columnwidth,0.19\columnwidth){\scriptsize {FS$_{M,o}$}}
\rput[tr](-0.25\columnwidth,0.1\columnwidth){\scriptsize {FS$_{\Gamma Z}$}}
\caption{(Color online) Fermi surface for $T>T_S$ (a) and $T<T_S$ (b) in the presence of a $d$-wave orbital order. As in Fig. 1 of the main text, the Fermi surfaces are derived from the 10 orbital model, plotted in the crystallographic BZ and visualized with the summed-color method where the absolute value of the overlap is mapped to the RGB value of the color on the surface as indicated by the color bar. The inset shows the shift in energy of the $d_{xz}$ band between $T$=170K and $T$=10K due to the effect of a $s+d$-wave orbital order. The $d_{xz}$ band (and similarly the $d_{yz}$ band) shifts in opposite directions at the $\Gamma$ and $X$-points.
}
\label{fig:FS_dwave}
\end{figure}
For $\Delta_{s}(T)>0$ and $\Delta_{d}(T)>0$, the $s$- and $d$-wave orbital components move the $d_{xz}/d_{yz}$ bands in opposite directions at the $\Gamma$ and $M$ points, respectively. Both $\Delta_{s}(T)$ and $\Delta_{d}(T)$ follow a mean field behavior with a maximum value of $25\,\text{meV}$ at $T=10\,\text{K})$. This leads to the hole band near $\Gamma$ being of primarily $d_{xz}$ ($d_{yz}$) character, and the inner electron band at M point of mainly $d_{yz}$ ($d_{xz}$) character as can be seen in the orbitally resolved band shown in Fig.\ref{fig:QOdwave}. This opposite shift of the bands can be seen more clearly from Fig.\ref{fig:FS_dwave} (inset) which displays the difference in band energy ($\Delta E = E_{xz}(T=170\,\text{K}))-E_{xz}(T=10\,\text{K})$) between the $d_{xz}$ band in orbital ordered state at $T=10\,\text{K})$ and no orbital ordering at $T=170\,\text{K})$. It shows clearly that the band shift due to orbital ordering moves in opposite directions at the $\Gamma$ and $X$ points, and therefore undergoes  a sign change along the $\Gamma-X$ direction (in the unfolded Brillouin zone). Unlike the Fermi surface of the case with pure $s$-wave orbital order (Fig.~\ref{fig:FS_no_SO}(b)), the opposite shift of the band (at $\Gamma$ and $X$ points) due to the $s+d$-wave orbital order leads to different orbital content between the hole and the inner electron Fermi surface pockets, as seen in the Fermi surface plot in Fig.\ref{fig:FS_dwave}(b).
The extremal orbits measurable by QO have also been calculated in the presence of $d$-wave orbital order. As seen in Fig.\ref{fig:QOdwave}, the Fermi surface areas agree well with experiments similar to previous results for purely $s$-wave orbital order. Additionally the Sommerfeld co-efficient calculated from the effective masses derived from the quantum oscilation calculation for $d$-wave orbital order gives a value of 4.1 mJ/mol-K$^2$.
\begin{figure}[bt]
\includegraphics[width=0.977\columnwidth]{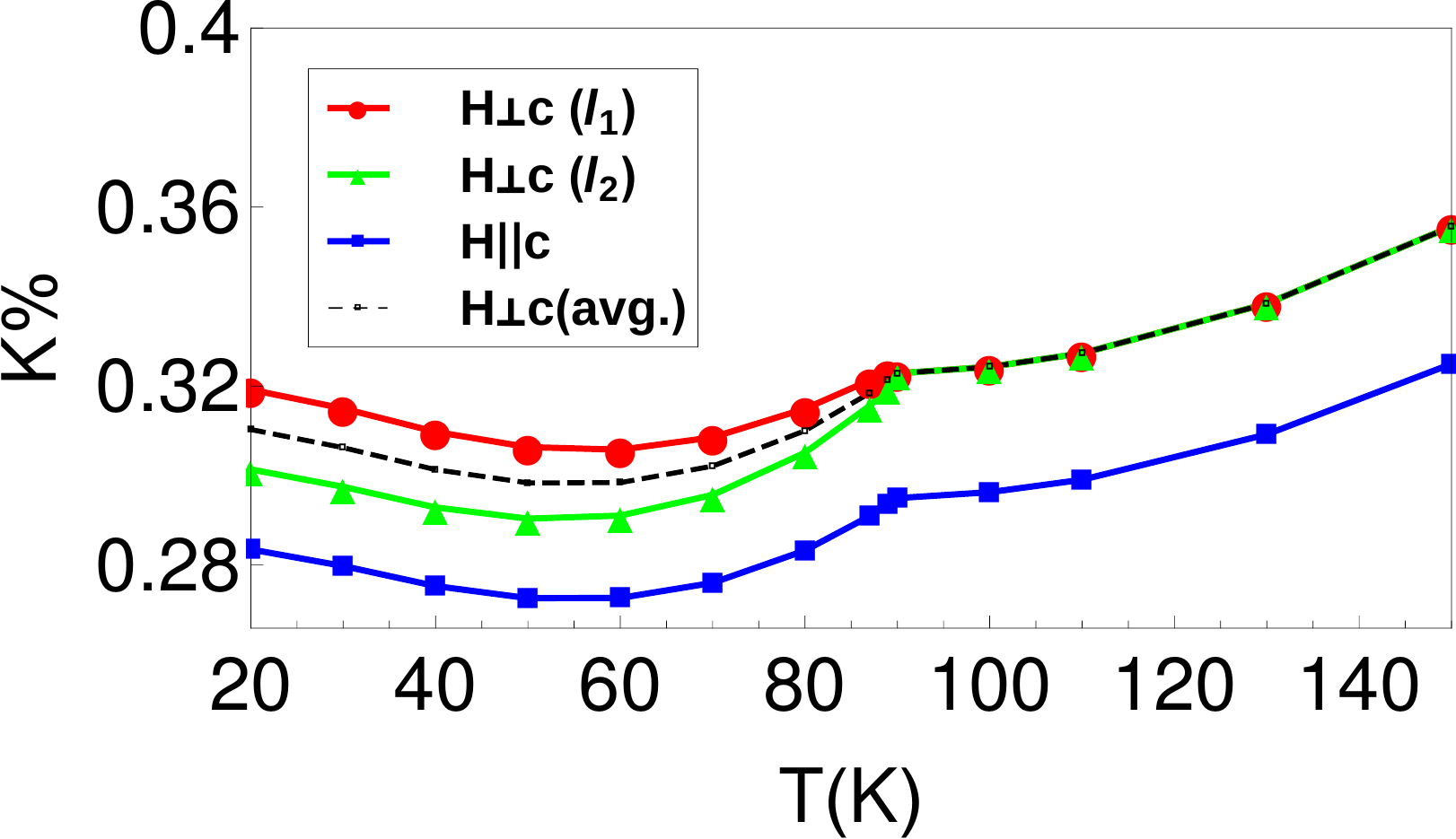}
\rput[tr](-0.95\columnwidth,0.56\columnwidth){(a)}\newline
\includegraphics[width=0.977\columnwidth]{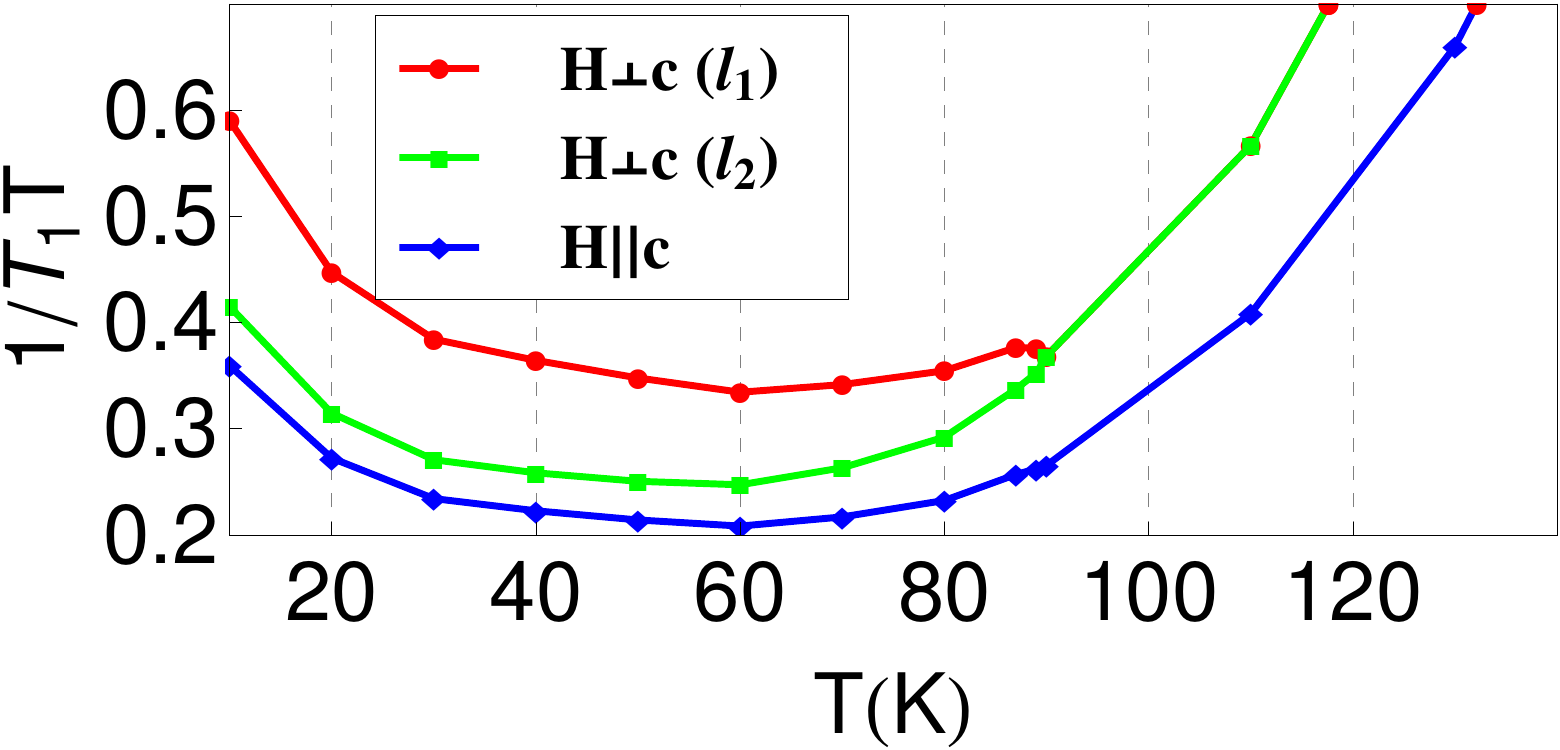}
\rput[tr](-0.95\columnwidth,0.52\columnwidth){(b)}\newline
\caption{(Color online) Plot showing the effects of $d$-wave orbital ordering on spin-lattice relaxation rate and NMR Knight shift.
(a) NMR Knight shift versus $T$. (b) Spin-lattice relaxation rate versus $T$. }
\label{fig:nmrdwave}
\end{figure}
Calculations of the NMR Knight shift and spin-lattice relaxation rate $1/T_1T$ in the presence of the $d$-wave orbitally ordered state are shown in Fig.\ref{fig:nmrdwave}. Evidently, general agreement with bulk experiments can also be obtained for this case.

\end{document}